\documentclass[aps,pre,twocolumn,amsmath,amssymb,nofootinbib,superscriptaddress,showpacs,longbibliography]{revtex4-2}

\usepackage{graphicx}
\usepackage{natbib}
\usepackage{latexsym}
\usepackage{amsmath, amssymb, amsthm}
\usepackage{mathtools}
\usepackage{xcolor}
\usepackage{comment}
\usepackage{wasysym}
\usepackage{slashed}

\usepackage[bottom]{footmisc}

\usepackage{microtype}
\usepackage{epigraph}
\usepackage{lipsum}

\usepackage{float}

\usepackage[normalem]{ulem}

\usepackage{appendix}

\usepackage{hyperref}

\newcommand{\km}{k_{\text{min}}}

\newcommand{\change}[1]{{\color{black}{#1}}}

\begin{document}

\title{Neighbor-induced damage percolation}

\author{Lorenzo Cirigliano}
\affiliation{Dipartimento di Fisica, Universit\`a \href{https://ror.org/02be6w209}{La Sapienza}, P.le
  A. Moro, 2, I-00185 Rome, Italy}


\author{Claudio Castellano}
\affiliation{\href{https://ror.org/05rcgef49}{Istituto dei Sistemi Complessi (ISC-CNR)}, Via dei Taurini
  19, I-00185 Rome, Italy}

\date{\today}

\begin{abstract}
We consider neighbor-induced damage percolation, a model describing systems
where the inactivation of some elements may damage their neighboring active
ones, making them unusable.
We present an exact solution for the size of the giant usable component
(GUC) and the giant damaged component (GDC) in uncorrelated random graphs.
We show that, even for strongly heterogeneous distributions,
the GUC always appears at a finite threshold and its formation is
characterized by homogeneous mean-field percolation critical
exponents.
The threshold is a nonmonotonic function of connectivity:
robustness is maximized by networks with finite optimal average degree.
We also show that, if the average degree is large enough,
a damaged phase appears, characterized by the existence of a GDC,
bounded by two distinct percolation transitions.
The birth and the dismantling of the GDC
are characterized by standard percolation critical exponents in
networks, except for the dismantling in scale-free networks
where new critical exponents are found.  Numerical simulations on
regular lattices in $D=2$ show that the existence of a GDC depends not
only on the spatial dimension but also on the lattice coordination number.
\end{abstract}


\maketitle

\change{
\section{Introduction}
}

Imagine a set of devices connected electrically to each other in a network,
as for example a power-grid.
A short-circuit in one of the nodes may cause damage also to other
nodes directly connected to it, making them unusable.
Analogously, a physical attack on a site of a technological network
may not only destroy it, but also disrupt activity in neighboring sites.
Similar types of phenomena may occur also in the context
of photonic quantum computing~\cite{lobl2024efficient}.
To model these systems and understand global connectivity
properties of both functioning and damaged nodes, we define a
\textit{neighbor-induced damage percolation} (NIDP) process,
in which the failure of a site may imply the damaging
of some of its neighbors.
\begin{figure}
\center
\includegraphics[width=0.45\textwidth]{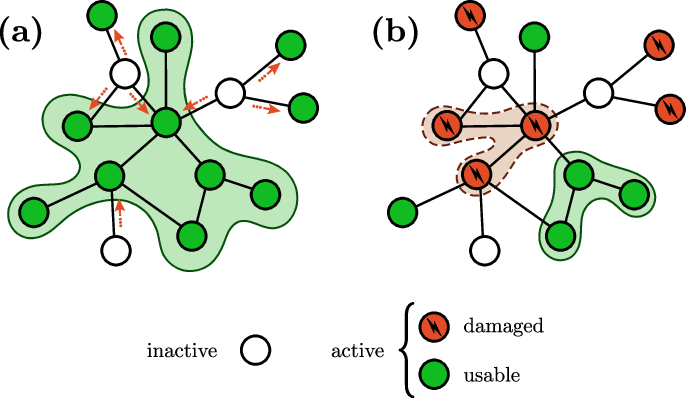}
    \caption{Schematic representation of neighbor-induced damage
      percolation for $\psi=1$. Empty nodes are inactive, filled nodes
      are active. (a) Configuration of active and inactive nodes. The
      standard largest component is highlighted in green. Inactive nodes induce
      damage on their active neighbors (red arrows). (b) Configuration
      of damaged (red) and usable (green) nodes. By contour lines we
      represent the largest usable component (continuous lines) and the largest damaged component (dashed
      lines). The effect of the neighbor-induced damage is visible by
      comparing the standard largest component and the largest usable component, the latter being
      significantly smaller than the former even if only a few nodes
      are inactive. 
      }
    \label{fig:schematic}
\end{figure}
Let us define the model precisely: in a network each node is active
with probability $\phi$ and inactive with probability $1-\phi$.
Inactive nodes have an effect on
their neighborhood: they may, independently with probability $\psi$,
\textit{damage} each of their active neighbors.
In this way, an active node is \textit{usable} if either it is
surrounded by active nodes only or it does not get damaged by any of
its inactive neighbors. Otherwise, an active node is
\textit{damaged} (see Fig.~\ref{fig:schematic}).
We are interested in
studying how usable connected components and damaged connected
components -- connected components made of usable and damaged nodes,
respectively -- behave as a function of $\phi$ and $\psi$ and in
particular the existence of a giant usable component (GUC) and of a
giant damaged component (GDC).
NIDP can also be seen as a model for enzymatic reactions of chemical
species: active nodes correspond to a given species S (substrate),
while inactive nodes represent an enzyme E. If an enzyme is in contact
with an S molecule, the latter can be transformed (with probability
$\psi$) into another species P (product), according to the reaction
$\text{E} + \text{S} \rightharpoonup \text{E} +
\text{P}$~\cite{nelson2017lehninger}. In such a framework the relevant
questions are if and how, depending on the initial density of enzymes,
extensive components of substrate (usable nodes) or products (damaged
nodes) are formed.

Despite its apparent simplicity, percolation theory still constitutes
a core research topic in statistical physics and complex
systems~\cite{araujo2014recent}.  Indeed, as the description of
real-world interacting systems has become more and more elaborate over
the years, percolation theory has evolved and adapted to such
complicated scenarios. Bootstrap percolation and
$k$-core~\cite{dorogovtsev2006kcore,goltsev2006kcore}, explosive
percolation~\cite{achlioptas2009explosive,dacosta2010explosive},
no-exclaves percolation~\cite{min2024noexclaves}, interdependent
percolation~\cite{buldyrev2010catastrophic,radicchi2015percolation},
percolation on multilayer and multiplex
networks~\cite{hackett2016bond,bianconi2016percolation,bianconi2017epidemic}
and on higher-order networks~\cite{sun2021higher,sun2023dynamic}, are
examples showing how recent research on percolation opened new
frontiers, unveiling a huge variety of phase transitions and
explaining complex behaviors.  Recently, new percolation models have
been proposed to face the increasing interest in quantum communication
networks where non-local effects are present. This is the case, for
instance, of extended-range
percolation~\cite{cirigliano2023extended,cirigliano2024general},
concurrence and
$\alpha$-percolation~\cite{meng2021concurrence,meng2023percolation}.
In these models, damage is local and the notion of connectivity
generalizes the standard one allowing for gaps in the paths connecting
active nodes.  Here we explore the opposite direction: models in which
the connectivity is local and the damage has non-local effect.  Some
percolation processes have already been investigated along this
line~\cite{kim2024shortest,goltsev2006kcore,shao2015percolation,yang2012network,yang2016observability,allard2014coexistence},
but NIDP differs from them. In particular in bootstrap
percolation~\cite{goltsev2006kcore}~\change{and localized attack
  percolation~\cite{shao2015percolation}} damage propagates~{\em
  iteratively}: here, instead, an active node damaged by one of its
inactive neighbors remains active, hence it does not damage any of its
active neighbors.  In the \textit{network observability
  problem}~\cite{yang2012network,yang2016observability}, some nodes in
a network are directly observed and they allow indirect observation of
their nearest neighbors.  In that case the focus is on the giant
component formed by directly or indirectly observable nodes, which in
our framework corresponds to the giant component made of inactive and
damaged nodes.  The distinction between usable and damaged nodes makes
NIDP markedly different from the network observability problem.
\change{The observability problem was generalized to depth-$L$
percolation in Ref.~\cite{allard2014coexistence}, where it was
pointed out the nontrivial coexistence of an observable and a
nonobservable extensive component.  In the specific case $\psi=1$
neighbor-induced damaged percolation is equivalent to depth-$L$
percolation for $L=1$ by mapping inactive, damaged and usable nodes
in NIDP to directly occupied, indirectly occupied and nonoccupied
nodes in depth-$L$ percolation, respectively. In this paper we
consider a more general model ($\psi \le 1$) and we perform a more detailed
analysis (in particular about damaged nodes) including critical
properties and low-dimensional systems.}

\change{
\section{Solution on locally tree-like random networks}

\subsection{The size of the giant components}

We start by deriving the equations for the size of the GUC and
of the GDC as a function of the control parameter $\phi$,
for arbitrary value of $0\leq \psi \leq 1$.

To simplify the equations, we introduce the parameter 
\begin{equation}
 \Phi = 1-\psi(1-\phi).
 \label{eq:Phi_def}
\end{equation}
Note that $\Phi$ ranges in $\phi \leq \Phi \leq 1$,
with $\Phi=\phi$ for the case $\psi=1$
and $\Phi=1$ for standard site percolation $\psi=0$.

An inactive node damages independently each of its active neighbors
with probability $\psi$. From the definition of usable node, we can
write an expression for the probability $U^{\infty}_{|k}$ that a
random node of degree $k$ is in the GUC. First of all, such a node
must be active. Assuming that this active node has $n$ active
neighbors and it does not get damaged by its $k-n$ inactive neighbors,
which happens with probability $(1-\psi)^{k-n}$, this node is in the
GUC if at least one of its active branches leads to the GUC, with
probability $[1-(1-u)^n]$.
Here $u$ is the probability that,
following a randomly chosen link that leads to an active
node, we reach the GUC along such branch.
Summing over all the possible values of
$n$, a random variable that follows a binomial distribution
$\mathcal{B}_n^k(\phi)={k \choose n}\phi^{k}(1-\phi)^{k-n}$, we get
\begin{align}
\nonumber
  U^{\infty}_{|k} &= \phi \sum_{n=0}^{k} \mathcal{B}_n^k(\phi) (1-\psi)^{k-n}[1-(1-u)^n] \\ \nonumber
  &= \phi \left[ \Phi^k - (\Phi - \phi u)^k \right].
\end{align}
Averaging this expression over the degree distribution $p_k$ we get
\begin{equation}
 U^{\infty} = \phi \left[g_0\big(\Phi \big)-g_0\big(\Phi - \phi u \big) \right],
 \label{eq:U_infty}
\end{equation}
where $g_0(z)=\sum_k p_k z^k$ is the generating function of the degree
distribution.

A similar computation can be carried out for the probability $u$.
Denoting by $u_{|r}$ the probability that a randomly chosen edge leading to an
active node $i$ of excess degree $r$ leads to the GUC, we can derive an
equation for its complementary probability $1-u_{|r}$ as follows.
Consider the case in which $n$ of
the $r$ residual neighbors of $i$ are active. Two mutually exclusive
scenarios can occur. The first possibility is that $i$ has been
damaged by one of its inactive neighbors, which happens with
probability $1-(1-\psi)^{r-n}$. If instead $i$ is usable, which
happens with probability $(1-\psi)^{r-n}$, then its active $n$
branches do not lead to the GUC with probability $(1-u)^n$. Summing
these two contributions and averaging over the binomial distribution
of the activation process over the $r$ residual neighbors we get
\begin{align}
\nonumber
  1-{u_{|r}} =& \sum_{n=0}^{r} \mathcal{B}_n^r(\phi)\left[1-(1-\psi)^{r-n} +(1-\psi)^{r-n}(1-u)^r \right]\\ \nonumber
 &= 1- \Phi^r +\left(\Phi - \phi u \right)^r,
\end{align}
from which, averaging over the excess degree distribution $q_r$ we get
\begin{equation}
u =  g_1 \big(\Phi \big) - g_1 \big(\Phi - \phi u \big),
\label{eq:u_psi}
\end{equation}
where $g_1(z) = \sum_r q_r z^r$.
Note that setting $\Phi=1$ we recover the standard
equations for percolation on uncorrelated random
graphs~\cite{newman2018networks}, while 
the equations for $\psi=1$ are recovered by setting $\Phi=\phi$.

%

We turn now our attention to the existence of a giant damaged
component. A damaged node is an active node which has been damaged by
at least one of its inactive neighbors. Denoting by $d$ the
probability that following a randomly chosen edge leading to an active
node we reach the GDC, we can write an equation for $D^{\infty}$, the
probability that a randomly chosen node belongs to the GDC. First of
all, this randomly chosen node must be active and one of its inactive
neighbors must have damaged it. Then, at least one of its active
neighbors must lead to the GDC. Conditioning on nodes of degree $k$ we
can write
\begin{align}
\nonumber
  D^{\infty}_{|k} &= \phi \sum_{n=0}^{k-1}\mathcal{B}_n^k(\phi)[1-(1-\psi)^{k-n}][1-(1-d )^n] \\ \nonumber
  &= \phi \left[1- \Phi^k + (\Phi-\phi d)^k + (1 - \phi d)^k \right],
\end{align}
where we extended the sum up to $k$ since the term with $n=k$,
corresponding to the scenario of an active node surrounded by active
neighbors, equals zero.  Averaging over $p_k$ we obtain
\begin{equation}
 D^{\infty}=\phi \left[1-g_0(\Phi)+g_0(\Phi-\phi d)-g_0(1-\phi d) \right].
 \label{eq:D_infty}
\end{equation}

In a similar way, we can write an equation for $d_{|r}$ conditioning
on nodes of excess degree $r$
\begin{align}
\nonumber
 1-d_{|r} &= \Phi^r + \sum_{n=0}^{r-1} \mathcal{B}_n^r(\phi) [1-(1-\psi)^{r-n}](1-d)^n \\ \nonumber
 &= \Phi^r + (1 - \phi d)^r - ( \Phi - \phi d)^r ,
\end{align}
and averaging over $q_r$ we get
\begin{equation}
d = 1-g_1(1-\phi d) + g_1(\Phi-\phi d) - g_1(\Phi) .
\label{eq:d_psi}
\end{equation}

As shown in Sec.~I of the Supplemental Material (SM),
}
the formalism allows us also to
write exact equations for the generating functions of the finite usable
and damaged component size distributions, $\pi_s^{U}$ and $\pi_{s}^{D}$,
respectively, from which we can compute the average size of the finite usable
and damaged components, denoted by $\langle s \rangle_{U}$ and
$\langle s \rangle_{D}$, respectively.

\begin{figure}
  \center
  \includegraphics[width=0.47\textwidth]{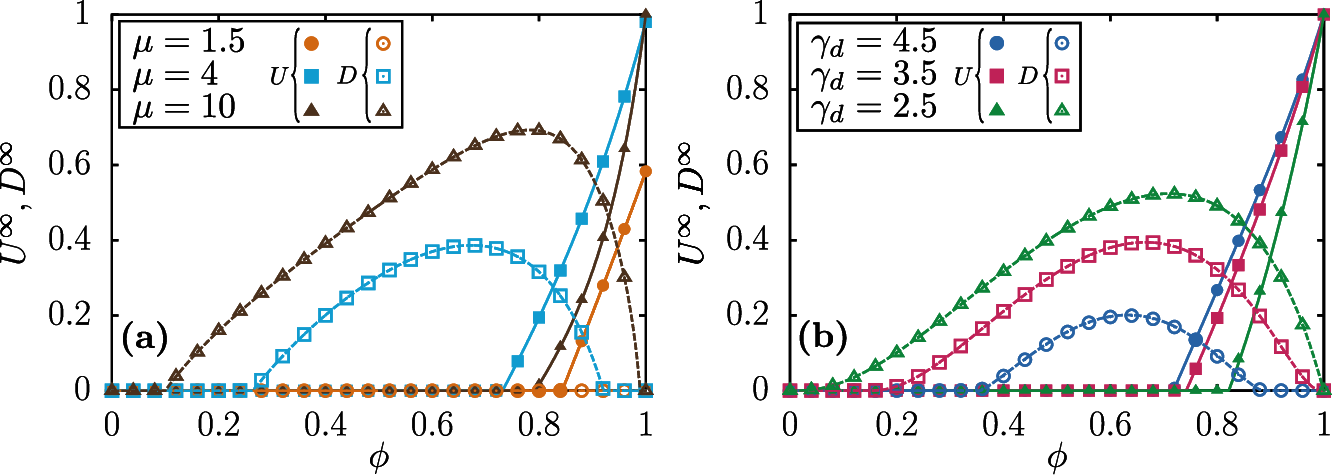}
  \caption{Order parameters $U^{\infty}$ [obtained from
      Eqs.~\eqref{eq:U_infty}-\eqref{eq:u_psi} (solid lines)]
    and $D^{\infty}$ [obtained from
      Eqs.~\eqref{eq:D_infty}-\eqref{eq:d_psi} (dashed
      lines)] as a function of $\phi$ under neighbor-induced damage
    with $\psi=1$, compared with numerical simulations (filled symbols
    for $U^{\infty}$, empty symbols for $D^{\infty}$) on networks with
    $N=10^6$ nodes, generated with the Uncorrelated Configuration
    Model~\cite{catanzaro2005generation}. Simulations are performed
    using a modified Depth-First Search algorithm, and results are
    averaged over $100$ independent realizations. (a) Erd\H{o}s- R\'enyi networks with
    $\mu=1.5$ (circles), $\mu=4$ (squares), and $\mu=10$
    (triangles). Note the absence of a GDC for $\mu=1.5<\mu_{*} \simeq 2.5804$.
    (b) Power-law networks with $\km=3$,
    $\gamma_d=4.5$, $\mu \simeq 3.65$
    (circles), $\gamma_d=3.2$, $\mu \simeq 4.30$ (squares), and
    $\gamma_d=2.5$, $\mu \simeq 7.64$ (triangles). Note that
    $\phi_{-}^{D}=0$ and $\phi_{+}^{D}=1$ for $\gamma_d=2.5<3$.}
    \label{fig:order_parameter}
\end{figure}

\change{
Focusing now on the case $\psi=1$,}
the solution of Eq.~\eqref{eq:u_psi}, plugged into
Eq.~\eqref{eq:U_infty}, gives the order parameter $U^{\infty}$
as a function of $\phi$. The solution of Eq.~\eqref{eq:d_psi},
plugged into Eq.~\eqref{eq:D_infty} gives the order parameter
$D^{\infty}$.  Both are shown 
in Fig.~\ref{fig:order_parameter} for
(a) Erd\H{o}s-R\'enyi (ER) and (b) power-law (PL) networks, with various average
degrees $\mu$~\change{and found in perfect agreement with numerical simulations}.

Fig.~\ref{fig:order_parameter} shows, as expected, the existence of a percolation
transition for usable nodes: for $\phi>\phi_c^U$ a GUC occupies a finite fraction of the
network. Remarkably, the threshold depends non-monotonically on the average degree $\mu$:
$\phi_c^U$ is minimal for an intermediate value of $\mu$, at odds
with standard percolation where increasing the average connectivity always reduces the threshold.
In addition, for power-law distributed networks the threshold is always
finite, also for scale-free networks with $\gamma_d \leq 3$, which have instead
zero threshold for standard percolation.
Concerning the GDC, an even richer phenomenology is observed in
Fig.~\ref{fig:order_parameter}.  For small $\mu$, $D^\infty=0$ for any
$\phi$: no GDC is ever possible (see panel (a) for $\mu=1.5$).  For
larger values of $\mu$, an extensive component of damaged nodes
appears in the interval $\phi_-^D<\phi<\phi_+^D$, the latter marking a
reentrant percolation transition.
In the case of scale-free networks with $2<\gamma_d \leq 3$, a GDC exists for any finite value o
f $\phi$, since $\phi_{-}^{D}=0$ and $\phi_{+}^{D}=1$.

\begin{figure}
\center
\includegraphics[width=0.48\textwidth]{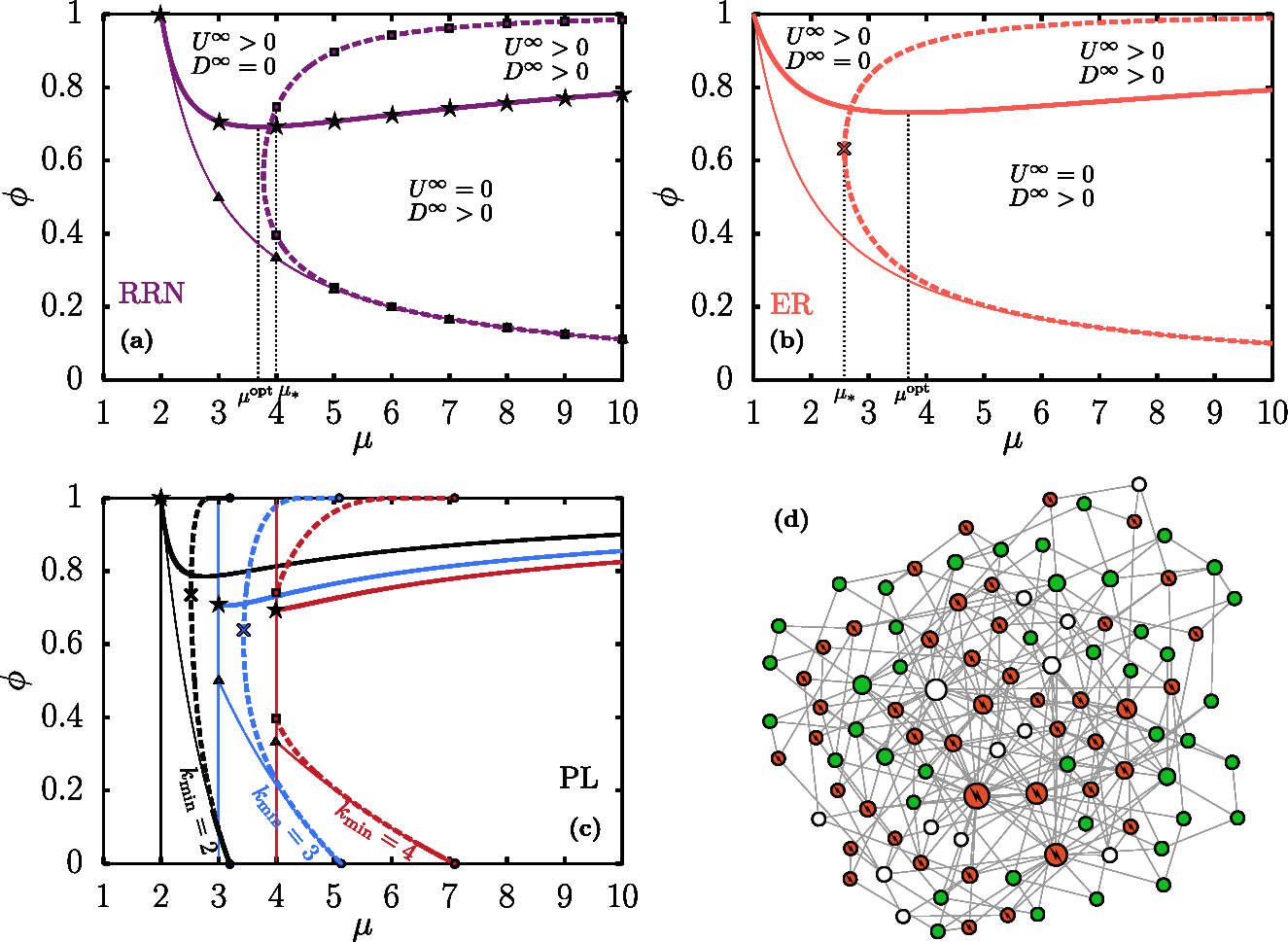}
\caption{NIDP thresholds $\phi_c^U$ (thick continuous lines) and
  $\phi_{\pm}^{D}$ (thick dashed lines) for $\psi=1$ obtained from
  Eq.~\eqref{eq:threshold_psi=1} and Eq.~\eqref{eq:damaged_threshold}
  as a function of the network average degree $\mu$, and compared with
  the standard site percolation thresholds $\phi_c^{\text{site}}=1/b$
  (thin lines). 
  (a) $\mu$-RRN networks. (b) ER networks. The vertical line
  indicates the position of the optimal average degree
  $\mu^{\text{opt}}=1+e$ and of the minimum required to observe a GDC
  $\mu_{*}$.  (c) Power-law networks with $p_k=
  k^{-\gamma_d}/\zeta(\gamma_d,k_{\text{min}})$ for $k \geq \km$ and
  $\gamma_d>2$, where $\zeta(\gamma_d,k_{\text{min}})$ is the Hurwitz
  zeta function~\cite{whittaker1996course} and $\km=2,3,4$.  For all
  networks a GDC phase exists for $\phi_{-}^{D}<\phi<\phi_{+}^{D}$,
  i.e., between the dashed lines, but only for $\mu>\mu_{*}$
  (indicated by the crosses). (d) A realization of NIDP on a small PL
  network of size $N=100$ with $\km=3$, $\gamma_d=2.5$. Nodes labels
  are as in Fig.\ref{fig:schematic}. The activation probability has
  ben set to $\phi=0.83$, in the phase where $U^{\infty}>0$,
  $D^{\infty}>0$. The largest usable and damaged components occupy a
  fraction $0.34$ and $0.36$ of the network, respectively. Note that
  network hubs are damaged.  }
\label{fig:threshold}
\end{figure}

\change{
\subsection{Threshold values}

The solution $u=0$ of the recursive
equation~\eqref{eq:u_psi} corresponds to a phase with no GUC.
Such a solution is stable if the first derivative of the
r.h.s. of Eq.~\eqref{eq:u_psi} evaluated at $u=0$ is smaller than
$1$. Hence the threshold condition is
\begin{equation}
 1=\phi_c^U g_1'(\Phi_c^U).
 \label{eq:critical_line_U}
\end{equation}
For $\phi>\phi_c^{U}$ another, stable, positive solution appears (see
the SM, Sec.~II.A, for details), corresponding to a phase with a GUC.

Concerning the existence of the giant damaged component, we can apply
the same stability criterion to Eq.~\eqref{eq:d_psi}. The solution
$d=0$ corresponds to a phase with no GDC, and a new stable solution
appears (see
the SM, Sec.~II.B, for details) when
\begin{equation}
 1=\phi_{\pm}^{D}\left[b-g_1'(\Phi_{\pm}^{D})\right],
 \label{eq:critical_line_D}
\end{equation}
where $\Phi_{\pm}^{D}=1-(1-\phi^{D}_{\pm})\psi$ and
$b=\langle k(k-1)\rangle/\langle k \rangle$ is the branching
factor~\cite{molloy1995critical,callaway2000network}.
We denote with
$\phi_{\pm}^{D}$ the solutions of Eq.~\eqref{eq:critical_line_D}
because we expect either no solution at all, two coincident
solutions, or two distinct solutions, according to the following
argument. Writing Eq.~\eqref{eq:critical_line_D} as $f(\phi,
\underline{\lambda} )=
\phi[b(\underline{\lambda})-g_1'(\Phi,\underline{\lambda})] - 1 = 0$,
where $\underline{\lambda}$ denotes the set of parameters on which the
degree distribution depends, we have $f(\phi)=-1$ for $\phi=0$ and for
$\phi=1$. Since $f$ is continuous and differentiable in $\phi$ for
$0<\phi<1$, there must exist at least one stationary point, say
$\phi_*$, in this interval. If $f(\phi_*)>0$, then the equation
$f(\phi^{D})=0$ admits at least two distinct solutions
$\phi_{\pm}^{D}$ in the interval $(0,1)$. We can conclude that,
depending on the choice of the parameters $\underline{\lambda}$, we
may have either no GDC, or a phase with a GDC for $0< \phi_{-}^{D} <
\phi < \phi_{+}^{D}<1$. Physically, this behavior is reasonable: if
$\phi$ is too small, there are too many inactive nodes to form a GDC,
while if $\phi$ is too close to $1$ almost everything is usable. From
this argument, it is safe to assume that there are at most two
solutions, corresponding to the birth and to the dismantling of the
GDC.

An important remark is in order here. Eq.~\eqref{eq:critical_line_D}
holds only for degree distributions with finite branching factor. In
fact, if $\gamma_d<3$ the solution $d=0$ is never stable for any
$0<\phi<1$ and any $\psi>0$, thus implying that the birth and
dismantling of the GDC take place only at $\phi_{-}^D=0$ and
$\phi_{+}^D=1$, respectively, for any $\psi>0$, while trivially no GDC exists
for $\psi=0$.

In the following, we discuss in detail the behavior of
Eq.~\eqref{eq:critical_line_U} and Eq.~\eqref{eq:critical_line_D}. We
begin consider for simplicity the case $\psi=1$, and then discuss what
happens for arbitrary $\psi<1$.

\subsubsection{The case $\psi=1$}

If the damaging of neighbors occurs with probability $1$,
Eq.~\eqref{eq:critical_line_U} becomes
\begin{equation}
 1=\phi_c^U g_1'(\phi_c^U).
 \label{eq:threshold_psi=1}
\end{equation}
}
A first nontrivial feature of NIDP can be deduced from
Eq.~\eqref{eq:threshold_psi=1}. In standard site percolation, networks
with broader degree distributions tend to be more robust. In
particular, the standard percolation threshold on random graphs is
$\phi_c^{\text{site}}=1/b$.
The solution of Eq.~\eqref{eq:threshold_psi=1} reveals instead a richer
behavior of $\phi_{c}^{U}$, depending on the network features and in
particular on the average degree $\mu=\langle k \rangle$.
For instance, for regular random networks with degree $\mu$ ($\mu$-RRN),
we have $g_0(z)=z^{\mu}$ and $g_1(z)=z^{\mu-1}$, with $b=\mu-1$, so
that $\phi_{c}^{U}=(1/b)^{1/b}$, which is non-monotonic in $b$, in
contrast to $\phi_c^{\text{site}}$, and has a minimum for
$\mu^{\text{opt}}=1+e$, as shown in
Fig.~\ref{fig:threshold}(a)\footnote{Note however that since for
RRN $\mu$ can take only discrete values, the actual minimum in this
case occurs for $\mu^{\text{opt}}=4$}.
For Erd\H{o}s- R\'enyi random graphs with average degree
$\mu=\langle k \rangle$, the branching factor is $b=\mu$, and
$g_0(z)=g_1(z)=e^{\mu(z-1)}$.
Eq.~\eqref{eq:threshold_psi=1} reads $1=\mu \phi_c^U e^{\mu(\phi_c^U-1)}$
which can be solved numerically, see
Fig.~\ref{fig:threshold}(b), while the position of the minimum of the
function $\phi_c^U(\mu)$ can be computed explicitly. Setting
$\phi_c'(\mu^{\text{opt}})=0$ we get the implicit equation
$\mu^{\text{opt}} = 1/[1-\phi_c(\mu^{\text{opt}})]$.  From the
criticality condition evaluated at $\mu^{\text{opt}}$ we then get
$\phi_c(\mu^{\text{opt}})=e/(1+e)$, hence also in this
case the minimum is reached for $\mu^{\text{opt}}=1+e$.
Finally, for networks with a power-law degree distribution
$p_k \sim k^{-\gamma_d}$ for $k \geq \km$ and $\gamma_d>2$,
results are shown in Fig.~\ref{fig:threshold}(c) for various $\km$.
For $\mu \to \km^{+}$, hence $\gamma_d \to \infty$, the networks
behave as $\mu-$RRN, compare symbols with Fig.~\ref{fig:threshold}(a).
As $\mu$ is increased ($\gamma_d$ is reduced) a non-monotonic behavior
is observed for $\km=2$ and $3$, while for $\km \geq 4$ the
minimum occurs for $\mu=\km$.
Remarkably, $\phi_c^{U}>0$ even for strongly heterogeneous networks
with $2 < \gamma_d \leq 3$, when $\phi_c^{\text{site}}=0$.
In particular, $\phi_c^U \to 1$ as $\gamma_d \to 2$ and $\mu \to \infty$.
For all these network classes we then conclude that robustness,
in terms of the onset of the GUC, is maximized by structures with finite
connectivity.
This is the result of the competition between two contrasting
effects: a large average degree implies many connections among
active nodes and thus large components made of active nodes;
but at the same time a large connectivity
allows inactive nodes to damage many of their active neighbors.

\change{Concerning the existence of a GDC, the condition for the threshold
Eq.~\eqref{eq:critical_line_D} becomes, for $\psi=1$,}
\begin{equation}
 1 = \phi_{\pm}^D[b-g_1'(\phi_{\pm}^D)].
 \label{eq:damaged_threshold}
\end{equation}
A finite GDC exists if this self-consistency equation admits solutions for
$0 \le \phi_{-}^D<\phi_{+}^{D} \le 1$.
Interestingly, this occurs only if the average connectivity is sufficiently
large $\mu>\mu_{*}$.
In particular (see Fig.~\ref{fig:threshold}(a))
for $\mu$-RRN there are no solutions if $\mu \le 3$
while there are two solutions $\phi_\pm^D(\phi)$ if $\mu \ge 4$: a finite
$D^\infty$ exists for values of $\phi$ between these thresholds.
The same happens for ER networks: a GCD exists only if
$\mu>\mu_{*}= \varphi + \ln(1+\varphi) \approx 2.5804 \ldots$, where
$\varphi=(1+\sqrt{5})/2$ is the golden ratio (see Eq.~\eqref{eq:mu_star_ER} below).
Note that, for large $b$, $\phi_{-}^{D} \to 1/b$, hence it tends to
coincide with the threshold of standard percolation.
For power-law degree-distributed networks $\mu_{*}$ is a quantity dependent on $\km$
(see Fig.~\ref{fig:threshold}\change{(c)}).
In such a case the thresholds are $0<\phi_-^D$ and $\phi_+^D<1$ as long as $\gamma_d>3$
while they are exactly 0 and 1 (i.e., a GDC exists for any finite $\phi$)
for \change{strongly heterogeneous} networks ($\gamma_d \le 3$).
From Fig.~\ref{fig:threshold} it is evident that the threshold
$\phi_{+}^D$ is not constrained by $\phi_c^U$: if, depending on the
network, $\phi_{+}^D<\phi_c^U$ then when $\phi$ is increased,
there is a phase where only the GDC exists followed by a phase where
no giant component exists, followed by a phase where only the GUC exists.
If instead $\phi_{+}^D>\phi_c^U$ there is a phase (for $\phi_c^U<\phi<\phi_+^D$)
with coexistence of both giant components (see Fig.~\ref{fig:threshold}(c,d))

\begin{figure}
\center
\includegraphics[width=0.485\textwidth]{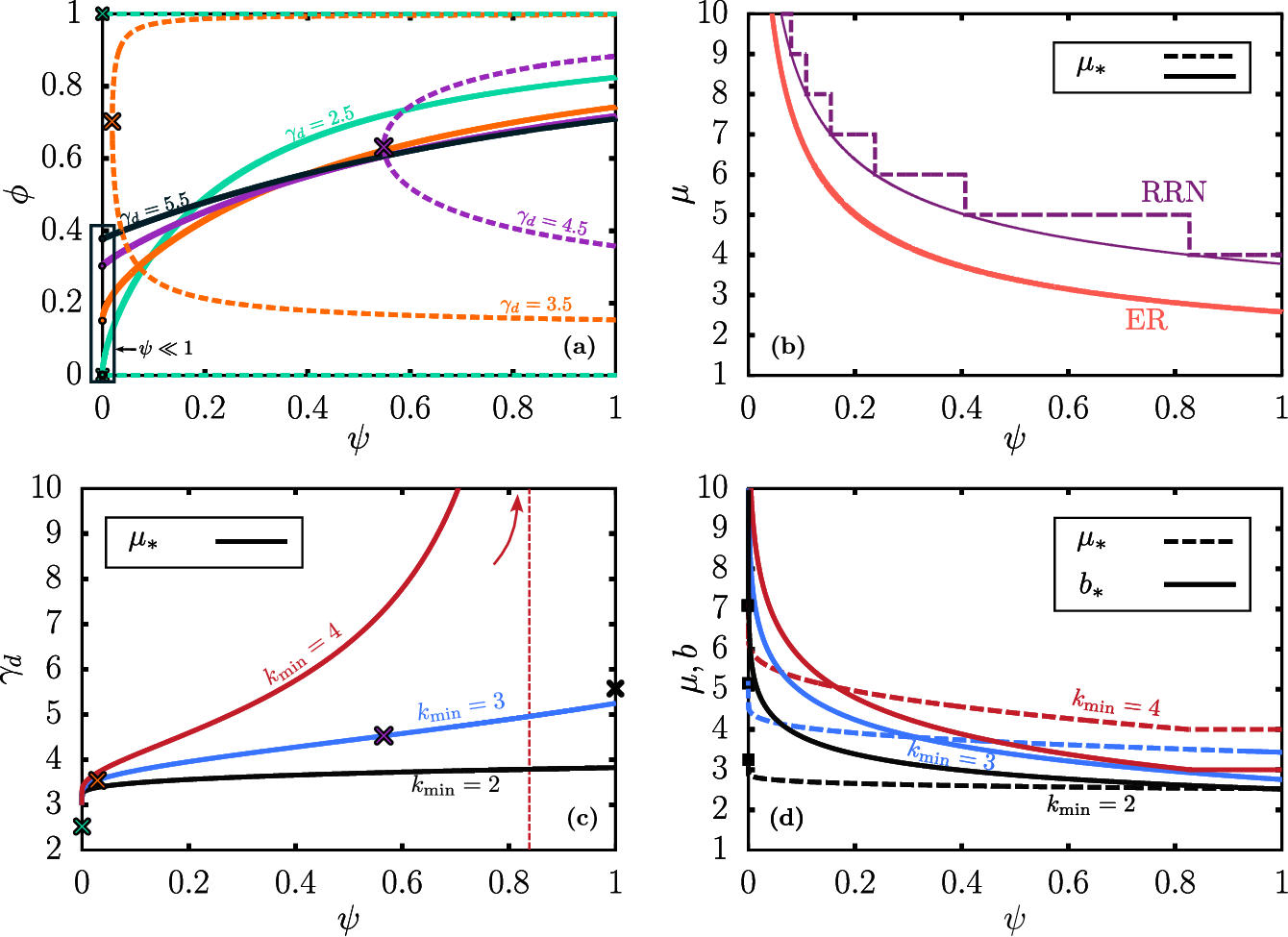}
    \caption{\change{(a) Phase diagram of NIDP for $0 \leq \psi \leq
        1$ in power-law networks with $\km=3$, and $\gamma_d=2.5$
        (green), $\gamma_d=3.5$ (orange), $\gamma_d=4.5$ (purple), and
        $\gamma_d=5.5$ (black). The thresholds $\phi_{c}^{U}$ solution
        of Eq.~\eqref{eq:critical_line_U} are represented by solid
        lines, the thresholds $\phi_{\pm}^{D}$ solutions of
        Eq.~\eqref{eq:critical_line_D} are represented by dashed
        lines. Circles at $\psi=0$ indicate the values of
        $\phi_{c}^{\text{site}}$: above the solid lines, a GUC exists,
        and inside the dashed regions a GDC exists. They may both
        exist at the same time. For the scaling of $\phi_c^{U}$ for $\psi
        \ll 1$ (inside the rectangle) see the SM, Sec.~II.A.
        The crosses mark the points at which the
        damaged phase disappears, i.e., the values of $\psi_{*}$ for
        the given values of $\gamma_{d}$ and $\km$, see panel
        (c). Note that a GDC always exists for $\gamma_d=2.5$ as soon
        as $\psi>0$, while no GDC exists for $\gamma_d=5.5$.(b)
        Numerical solution of Eqs.~\eqref{eq:curve_1} and
        \eqref{eq:curve_2} with $\lambda_{*}=\mu_{*}$ for RRN (thin
        line, dashed line represents the ceiling of such a solution)
        and the exact expression for $\mu_*$,
        Eq.~\eqref{eq:mu_star_ER}, for ER (continuous line).
        They both diverge for $\psi \to 0$. (c) Numerical solution of
        Eqs.~\eqref{eq:curve_1} and \eqref{eq:curve_2} with
        $\lambda_{*}=\gamma_{d*}$ for PL networks with various
        $\km$.
        Curves represent, given $\psi$, the maximum
        $\gamma_{d}$ for the existence of a damaged
        phase. Equivalently, given $\gamma_{d}$, they represent the
        minimum $\psi$ required to observe a damaged phase. Compare
        the crosses with the crosses in panel (a). All curves tend to
        $\gamma_{d*}=3$ for $\psi \to 0$, since a GDC always exists
        for $\gamma_d<3$ for any finite $\psi>0$. (d) Same as in panel
        (c), but using $\mu_{*}=\mu(\gamma_{d*},\km)$ (dashed lines)
        and $b_{*}=b(\gamma_{d*},\km)$ (continuous lines) instead of
        $\gamma_{d*}$.}}
    \label{fig:phase_diagram}
\end{figure}

\change{
\subsubsection{The case $\psi<1$}

We now analyze how $\psi$ comes into play. We begin by considering the
formation of the giant usable component. The solution of
Eq.~\eqref{eq:critical_line_U} for PL networks with $\km=3$ and
various $\gamma_d$ is reported in Fig.~\ref{fig:phase_diagram}(a). The
non-monotonic behavior of $\phi_{c}^{U}$ as a function of $\gamma_d$
at fixed $\psi$ is found for any finite $\psi$. Note that $\Phi<1$ if
$\psi>0$ and $\phi<1$. In this case, the singularities of the
generating functions typically encountered when dealing with power-law
distributions are always avoided, since $g_1(z)$ and all its
derivatives are finite for a fixed $z<1$, as they may diverge only for
$z \to 1^{-}$. Hence Eq.~\eqref{eq:critical_line_U} is well-defined
for any degree distribution. This implies, as can be seen from
Fig.~\ref{fig:phase_diagram}(a), that $\phi_c^{U}>0$ even for strongly
heterogeneous networks as soon as $\psi>0$.

We now consider the percolation transition of the GDC.  As it will be
shown below, reducing $\psi$ has, in practice, an effect analogous to
reducing the average degree of the network: if a GDC is present for
$\psi=1$, then reducing $\psi$ below a threshold value $\psi_{*}$
leads to the disappearance of the GDC, unless the network is strongly
heterogeneous ($\gamma_d \le 3$).  A GDC can exist only if
Eq.~\eqref{eq:critical_line_D} admits two distinct solutions. This
happens or not depending on the parameters $\underline{\lambda}$. The
criticality condition $f(\phi_{\pm}^{D}, \underline{\lambda})=0$
ensures the existence of two distinct solutions as soon as
$f(\phi_*,\underline{\lambda})>0$, where $\phi_*$ is the stationary
point of $f$.
When $f(\phi_*, \underline{\lambda})=0$, we have
$\phi_{*}=\phi_{-}=\phi_{+}$ and the two solutions coincide. Denoting
with $\underline{\lambda}_{*}$ the values of the parameters for which
this happens, we have two conditions for $\phi_*$ and
$\underline{\lambda}_{*}$
\begin{align}
\label{eq:curve_1}
f(\phi_*, \underline{\lambda}_{*}) &= \phi_{*} g_{1}'(\Phi_{*},\underline{\lambda}_{*}) - \phi_{*}b(\underline{\lambda}_{*}) + 1 = 0,\\
\label{eq:curve_2}
\partial_{\phi}f(\phi_*, \underline{\lambda}_{*}) &= g_{1}'(\Phi_{*},\underline{\lambda}_{*}) + \psi \phi_{*} g_{1}''(\Phi_{*},\underline{\lambda}_{*}) - b(\underline{\lambda}_{*}) = 0.
\end{align}
In general, the degree distribution may depend on an arbitrary set of
parameters $\underline{\lambda}$. Here we consider the cases of random
regular network and of Erd\H{o}s-R\'enyi networks, in which there is
only one parameter, the average degree $\mu$ (equivalently, the
branching factor $b=\mu-1$), and the case of power-law networks with
$p_k=k^{-\gamma_d}/\zeta(\gamma_d,\km)$, in which there are two
parameters $\gamma_d$ and $\km$. For fixed $\km$, we can use then
$\gamma_{d}$ as the only parameter. Hence for fixed $\psi$ the
solution of Eqs.~\eqref{eq:curve_1},\eqref{eq:curve_2} represents an
intersection between two curves in the $\phi-\lambda$ plane. At fixed
$\psi$, for homogeneous distributions the value $\mu_{*}$ corresponds
to the minimum average degree required to observe a damaged phase, see
Fig.~\ref{fig:phase_diagram}(b), while for PL networks the solution
$\gamma_{d*}$ is the maximum value $\gamma_{d}$ can take in order to
observe a damaged phase, see Fig.~\ref{fig:phase_diagram}(c). If
we fix instead $\mu$ or $\gamma_{d}$, the solution of
Eqs.~\eqref{eq:curve_1},\eqref{eq:curve_2} gives us the minimum value
$\psi_{*}$ of the damaging probability in order to observe a GDC. Even
if in general Eqs.~\eqref{eq:curve_1},\eqref{eq:curve_2} must be
solved numerically, for ER networks we get an explicit expression for
$\mu_{*}$ (see the SM, Sec.~II.C)
\begin{equation}
\mu_{*} = \varphi(\psi) + \frac{\ln \left[1+\psi \varphi(\psi)\right]}{\psi},
\label{eq:mu_star_ER}
\end{equation}
where $\varphi(\psi)=(1+\sqrt{1+4/\psi})/2$.

In Fig.~\ref{fig:phase_diagram}(d), for PL networks, we also plot
$\mu_{*}=\mu(\gamma_{d*},\km)$ and $b_{*}=b(\gamma_{d*},\km)$ as
functions of $\psi$, in order to compare with the analogous results
for homogeneous degree distributions in
Fig.~\ref{fig:phase_diagram}(b). Contrary to what happens with the
usable phase transition, in this case the relevant parameter is the
branching factor $b$ rather than $\mu$. In other words, the network
heterogeneity, in particular when $2<\gamma_d \leq 3$, plays a crucial
role.  }

\change{\subsection{Critical properties}}

\begin{figure}
  \center \includegraphics[width=0.485\textwidth]{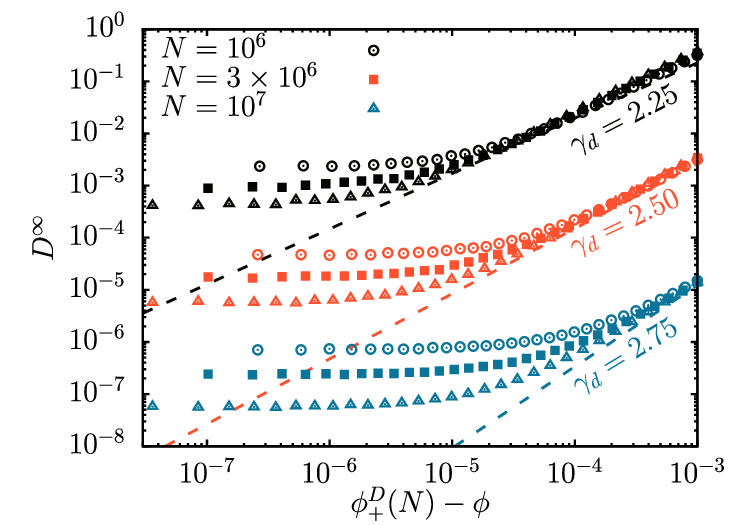}
  \caption{\change{Fraction of nodes in the largest damaged component,
      from numerical simulations of the NIDP process for $\psi=1$ on power-law networks with
      various values of $\gamma_d$, and sizes $N=10^6$ (circles), $N=3\times 10^6$ (squares),
      $N=10^7$ (triangles).
      The structural cutoff $k_c=N^{1/2}$ has been used~\cite{catanzaro2005generation}.
      $\phi_+^{D}(N)$ has been estimated as the position of the peak of the average size
      of finite damaged components. Results are averaged over 1000 independent realizations.
      Dashed lines are the theoretical prediction $\beta_{+}^{D}=1+(\gamma_d-2)^2$.
      The data for $\gamma_d=2.25$ and $\gamma_d=2.75$ are shifted upward an downward,
      respectively, for a better visualization. 
}}
  \label{fig:confirmation}
\end{figure}

It is also possible to fully analyze the critical properties of the
NIDP transitions (see the SM, Sec.~III, for detailed computations). \change{The
  role of $\psi$ is not relevant for the determination of the critical
  exponents, provided $\psi >0$.}
Concerning the formation of the GUC,
we can compute the critical exponent $\beta^{U}$, defined by
$U^{\infty}/\phi \sim (\phi-\phi_c^{U})^{\beta^{U}}$, and
$\gamma^{U}$, defined by $\langle s \rangle _{U}-1 \sim |\phi -
\phi_c^{U}|^{-\gamma^{U}}$, where $\langle s \rangle_{U}$ is the
average size of finite usable components.  We find $\beta^{U}=1$ and
$\gamma^{U}=1$, the homogeneous mean-field critical
exponents~\cite{cirigliano2024scaling} for standard percolation.
\change{While this is trivial for homogeneous networks and even for
  power-laws with $\gamma_d>4$, it is remarkable that these} values
hold for arbitrary degree distributions, i.e., even for power-laws
with $\gamma_d<4$, where instead exponents for standard percolation
depend on $\gamma_d$~\cite{cirigliano2024scaling}.  The origin of this
change can be understood by considering that the neighbor-induced
damaging process implies that hubs in NIDP are typically damaged. This
produces a finite cutoff in the degree distribution of usable nodes on
any network, at odds with what happens with uniform damaging (standard
percolation), which does not modify the nature of the large-$k$ tail
of the degree distribution.  Concerning the damaged phase transition,
the critical exponents $\beta^{D}_{\pm}$ are defined by
$D^{\infty}/\phi \sim [\mp (\phi-\phi_{\pm}^{D})]^{\beta^{D}_{\pm}}$,
while $\gamma_{\pm}^{D}$ are defined by $\langle s \rangle _{U}-1 \sim
|\phi -\phi_{\pm}^{D}|^{-\gamma^{D}_{\pm}}$, where $\langle s \rangle_{D}$
is the average size of finite damaged components.  We
find standard homogeneous mean-field critical exponents
($\beta=1$, $\gamma=1$) for both transitions in homogeneous networks
and power-laws with $\gamma_d>4$, and heterogeneous mean-field
values $\beta=1/(\gamma_d-3)$, $\gamma=1$, for both transitions in
heterogeneous networks with $3<\gamma_d<4$.  \change{Both for
  homogeneous networks and for power-law networks with $\gamma_d>3$}
the thresholds $\phi_{-}^{D}>0$ and $\phi_{+}^{D}<1$ correspond to
\change{usual} \textit{critical} phase transitions, since $\langle s
\rangle_{D}$ diverges in both cases.  The exponents are the same of
standard percolation because the large-$k$ tail of the degree
distribution of damaged nodes is the same.  When $\gamma_d<3$,
we find, close to $\phi_{-}^{D}=0$, the same exponents
$\beta_{-}^{D}=1/(3-\gamma_d)$ and $\gamma_{-}^{D}=-1$ of standard
percolation: the formation of a GDC is analogous to the formation of
the standard GC.  We find instead, close to $\phi_{+}^{D}=1$, new
exponents $\beta_{+}^{D}=1+(\gamma_d-2)^2$ and
$\gamma_{+}^{D}=-1+2(\gamma_d-2)(3-\gamma_d)$, indicating that the
dismantling of the GDC is determined by a qualitatively different
mechanism. \change{See Fig.~\ref{fig:confirmation} for a numerical
  confirmation of the value of $\beta_{+}^{D}$, and the SM, Sec.~III. for the
  numerical solution of the recursive equations for arbitrary large
  cutoff $k_c$.}  \change{Note that $\gamma_{-}^{D}<0$
  and $\gamma_{+}^{D}<0$, hence in both cases $\langle s \rangle_{D}$ does not
  diverge at the transition.}

\change{\section{Two-dimensional lattices}} Finally, we 
considered NIDP in regular lattices.
In particular, in $D=2$ we find that \change{for $\psi=1$} a GDC does not
exist in square lattices \change{(with coordination number $\mu=4$)}
while it exists in triangular lattices \change{($\mu=6$)}, see
Fig.~\ref{fig:lattices}. Hence, also in this case we find that the
existence of the GDC is possible only for sufficiently large
connectivity.  Finite-size scaling (see SM, Sec.~IV) indicates that
$\phi_{+}^{D}\leq \phi_{c}^{U}$, hence there is no coexistence between
a GUC and a GDC in the infinite-size limit\change{ for triangular
  lattices in $D=2$}.  \change{If the damaging of neighbors happens with
  probability $\psi<1$, the picture is qualitatively the same as for
  random graphs. Decreasing the value of $\psi$, the threshold
  $\phi_c^{U}$ tends to the percolation threshold of standard 
  site percolation in $D=2$.

  Concerning the damaged phase transition, we show
  in Fig.~\ref{fig:lattices2} that the existence of the GDC in
  the triangular lattice occurs only for sufficiently large values of $\psi$.
  In panel (a), for $\psi=0.58$, the relative
  size of the largest damaged component goes to zero as $N$
  increases. For small sizes, two peaks can be observed in $\langle s
  \rangle_{D}$; however, increasing the system size these two peaks
  disappear and only one peak is left: No GDC exists in the
  thermodynamic limit. In panel (b), for $\psi=0.6$, two clear peaks are
  present in the susceptibility (bottom panel) and grow with the
  system size, corresponding to the birth and the dismantling of a GDC.
  We can conclude that $0.58\leq \psi^{*} \leq 0.6$ for triangular lattices.

In all cases, a detailed finite-size scaling analysis (see SM, Sec.~IV)
demonstrates, as expected, that both the usable and the damaged
transitions belong to the same universality class of standard site
percolation in $D=2$.
\begin{figure}
  \center \includegraphics[width=0.47\textwidth]{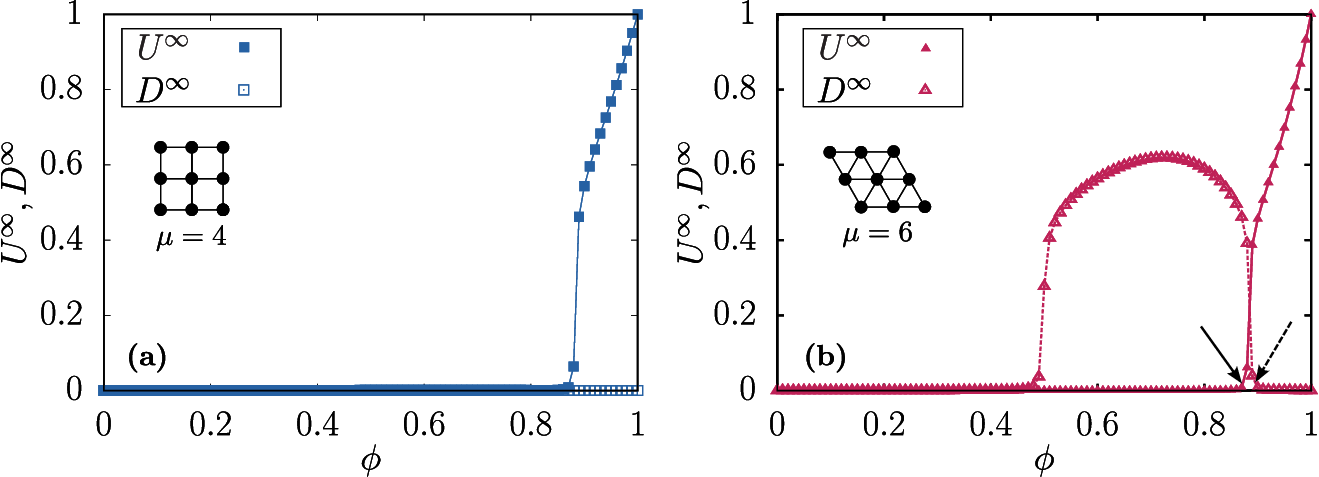}
  \caption{Numerical simulations of NIDP for $\psi=1$ in $D=2$ on (a) square
    lattices, (b) triangular lattices, of size $L=10^3$,
    $N=10^6$. Results are averaged over \change{1000 independent
    realizations}.
    Lines are a guide to the eye.
  }
  \label{fig:lattices}
\end{figure}
\begin{figure}
  \center \includegraphics[width=0.47\textwidth]{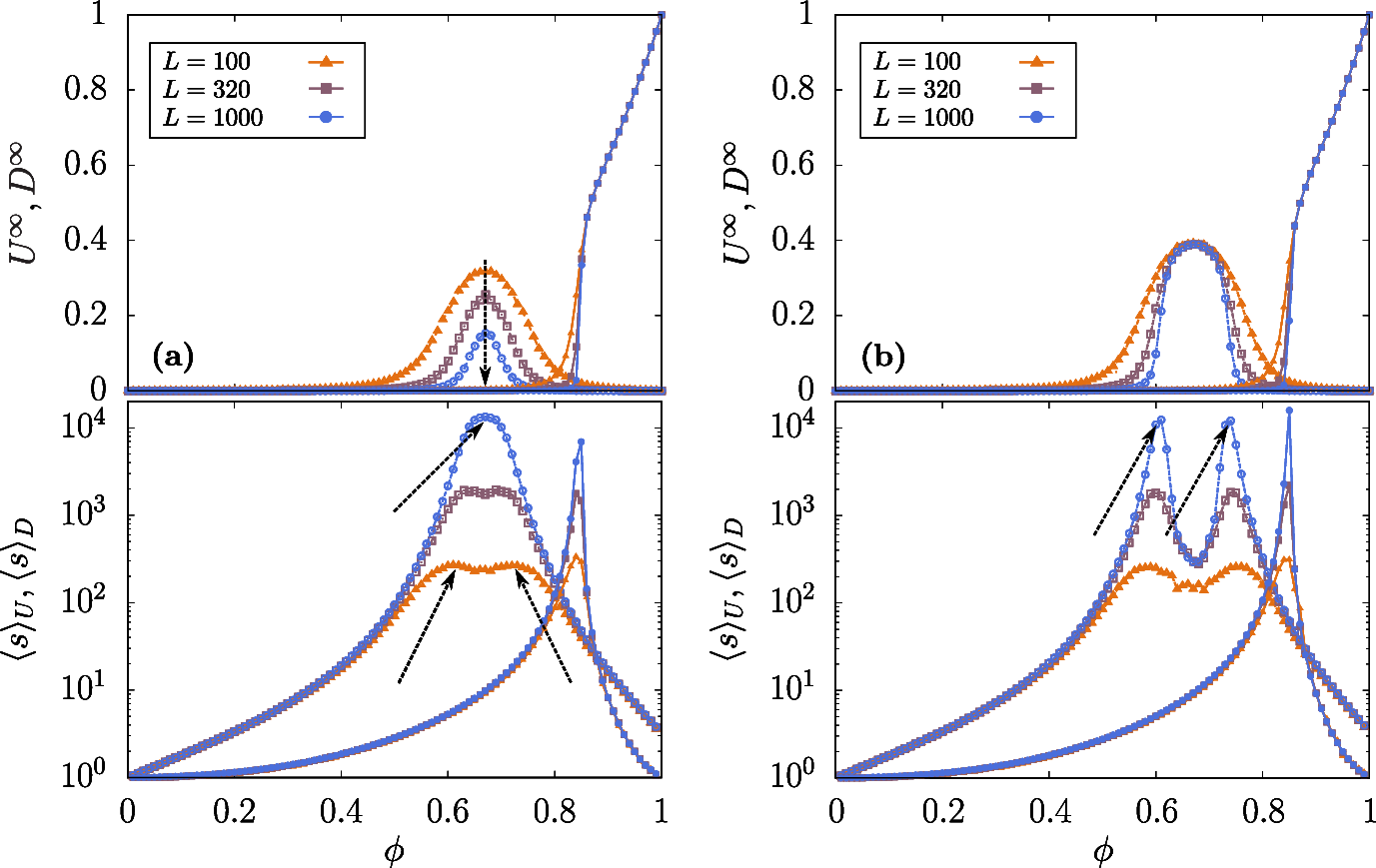}
  \caption{\change{Numerical simulations of NIDP on triangular
      lattices in $D=2$ for different values of $\psi$ and various
      sizes. Results for the observables related to the usable and the
      damaged transitions are represented with full and empty symbols,
      respectively, and they are averaged over 1000 independent realizations. Lines are a guide to the eye. Top: the order
      parameters $U^{\infty}$ and $D^{\infty}$. Bottom: the average
      usable and damaged cluster sizes $\langle s \rangle_{U}$ and
      $\langle s \rangle_{D}$. (a) $\psi=0.58$. (b) $\psi=0.60$.
      Note that in (a), no GDC is present in the infinite-size limit,
      while in (b) a GDC is present, as two peaks in $\langle s \rangle_D$ are
      clearly visible as $N$ increases.}}
  \label{fig:lattices2}
\end{figure}
}

\change{\section{Conclusions}}

In this \change{manuscript, we introduced and studied} a percolation
model called \textit{neighbor-induced damage percolation} (NIDP), in
which inactive nodes may damage their active neighbors \change{with
  probability $\psi$}, making them unusable. This process defines
three distinct classes of nodes: inactive, active usable, and active
damaged.  The exact solution of NIDP in uncorrelated random graphs in
the infinite size limit reveals a rich phase diagram \change{with} both
usable and the damaged percolation transitions, triggered by different
physical mechanisms and characterized by different critical
properties. In particular, a giant usable component exists only above
a threshold value $\phi_c^U$ of the fraction of active nodes, which is,
for any network class, a nonmonotonic function of the average
connectivity $\mu$, thus determining the existence of an optimally
robust network for a finite value of $\mu$.  For the existence of a
giant damaged component it is instead necessary a sufficiently large
average connectivity $\mu>\mu_*$. As a function of $\phi $ the GDC is observed
only in an interval: no GDC is possible when there are too few or too
many active nodes. \change{A reduction of the damaging probability $\psi$
favors the formation of a GUC, while it tends to shrink the interval
of $\phi$ values for which a GDC exists, which disappears altogether for $\psi<\psi_*$.
  Remarkably, for strongly heterogeneous networks a GDC
  always exists for $0<\phi<1$ as soon as $\psi>0$. We also found new
  critical exponents governing the dismantling of the GDC for $\phi
  \to \phi_{+}^D=1$, which do not correspond to the standard
  heterogeneous mean-field universality class.}

The same phenomenology is observed, by means of numerical simulations,
in $2$-dimensional lattices, where the existence of a GDC
strongly depends on microscopic details, in particular the lattice
coordination number.
Further investigations are needed to understand more deeply and
quantitatively this behavior in finite-dimensional systems.
The consideration of neighbor-induced damage percolation in
real-world networks, such as power-grids or road networks
and metabolic or ecological networks, is an interesting avenue
for future research, requiring possibly the development of a
message-passing approach to the problem.
In addition, many questions are open about optimal properties of NIDP.
Given a fixed average degree $\mu$ what is the topology
that maximizes (or minimizes) the interval of $\phi$ values for which
a GDC exists? Or the size of the GDC?
Given a network, what strategies to place inactive nodes lead to
optimal birth or dismantling of the giant damaged component?
Finally, the interpretation of NIDP in terms of enzymatic reactions sets the basis for a more realistic generalizations of the model to the case $\text{E}+\text{S} \xrightleftharpoons[\kappa_2]{\kappa_1} \text{E} + \text{P}$, where $\kappa_{1}, \kappa_{2}$ are reaction rates. The exploration of such an out-of-equilibrium system requires further studies.

\bibliography{NIDP_BIB}

\end{document}


\title{Supplemental Material for ``Neighbor-induced damage percolation''}

\author{Lorenzo Cirigliano}
\affiliation{Dipartimento di Fisica Universit\`a ``\href{https://ror.org/02be6w209}{La Sapienza}”, P.le
  A. Moro, 2, I-00185 Rome, Italy}


\author{Claudio Castellano}
\affiliation{\href{https://ror.org/05rcgef49}{Istituto dei Sistemi Complessi (ISC-CNR)}, Via dei Taurini
  19, I-00185 Rome, Italy}

\date{\today}

\begin{abstract}

\end{abstract}

\maketitle
In this Supplemental Material we derive the recursive equations for the generating functions of the cluster size distribution for both the usable and damaged components, from which we get the exact expressions for the average size of finite usable and damaged components, compared with numerical simulations.

We provide some additional details on the study of the existence of a damaged component for arbitrary $\psi$.

We discuss in detail the behavior, in uncorrelated random graphs, of the usable and damaged observables close to the thresholds, computing the values of the critical exponents for both the usable and the damaged transitions.

Finally, we provide some additional numerical studies, with a detailed finite-size scaling analysis, of neighbor-induced damage percolation in two-dimensional square and triangular lattices.

\section{Equations for the components size distributions}
\label{appendix:recursive}

\subsection{Finite usable components}

We turn now our attention to finite components. Following the
standard literature~\cite{christensen2005complexity,stauffer2018introduction},
let us denote by $\mathcal{N}^{U}_s$ the number of finite usable components. Then
$n_s^U=\mathcal{N}_s^U/(\phi N)$ is the density of usable components per
active site, and $\pi_s^U=sn_s^U$ is the probability that the size of a
finite usable component to which a randomly chosen active node belongs is $s$.

Defining $\rho_{s}^U$ as the probability that following a
random link leading to an active node we reach a finite usable
component of size $s$, for a locally tree-like network we can write
\begin{equation}
 \pi_{s}^U = \sum_{k}p_k \sum_{s_1,\dots,s_k} \delta_{s_1+\dots+s_k,s-1}\sum_{n=0}^{k}{k \choose n}\prod_{i=1}^n(\phi \rho_{s_i}^U)\prod_{j=n+1}^{k}[(1-\phi)(1-\psi) \delta_{s_j,0}],
\end{equation}
where the factor $(1-\phi)(1-\psi)$ is the probability that the
inactive neighbors do not corrupt the focal node, and $\delta_{s,0}$ appears
because inactive neighbors lead only to finite usable components of
size $0$.  Multiplying by $z^s$ and summing over $s$ we can get the
expression for the generating function $H_0^U(z)=\sum_{s}\pi_s^U z^s$
of the usable components size distribution
\begin{align}
\nonumber
 H_0^U(z)&= \sum_{k}p_k\sum_{s \geq 0}z^s\sum_{s_1,\dots,s_k} \delta_{s_1+\dots+s_k,s-1}\sum_{n=0}^{k}{k \choose n}\prod_{i=1}^n(\phi \rho_{s_i}^U)\prod_{j=n+1}^{k}[(1-\phi)(1-\psi) \delta_{s_j,0}]\\ \nonumber
 &=z \sum_{k} p_k \sum_{n=0}^{k}{k \choose n}\prod_{i=1}^n(\phi \sum_{s_1 \geq 0}\rho_{s_i}^U z^{s_i})\prod_{j=n+1}^{k}[(1-\phi)(1-\psi) \sum_{s_j\geq 0}\delta_{s_j,0}z^{s_j}]\\ \nonumber
 &=z g_0 \left((1-\phi)(1-\psi) + \phi H_1^U(z) \right),\\
 &=zg_0 \left(\Phi - \phi + \phi H_1^U(z) \right),
 \label{eq:H_0}
\end{align}
where $\Phi = 1-\psi(1-\psi)$.

If we follow a link leading to an active node of excess degree $r$, we
can either arrive to a damaged node, or to an usable node. The first
scenario corresponds to the probability of reaching a damaged node
$\rho_{0}^U = 1 - g_1(\Phi)$, while for $s>0$ we can write
\begin{widetext}
 \begin{align}
  \rho_{s}^U = \sum_{r} q_r \sum_{s_1,\dots,s_r} \delta_{s_1+\dots+s_r,s-1}\sum_{n=0}^{r}{r \choose n}\prod_{i=1}^n(\phi \rho_{s_i}^U)\prod_{j=n+1}^{r}[(1-\phi)(1-\psi) \delta_{s_j,0}].
 \end{align}
\end{widetext}

Multiplying by $z^s$ and summing over $s$, following the same
computation performed some lines above, we get
\begin{equation}
 H_1^U(z)=1-g_1(\Phi)+zg_1\left(\Phi - \phi + \phi H_1^U(z) \right),
 \label{eq:H_1}
\end{equation}

Note that $H^{U}_1(1)=1-u$ is the probability that following a link ending
in an active node we reach a finite usable component, hence we do not
reach the GUC. $H^{U}_0(1)$ is the probability that a randomly chosen
active node belongs to a finite usable component of any size. A
randomly chosen node belongs to the GUC with probability $U^{\infty}$
by definition. If this is not the case, it can either be inactive with
probability $1-\phi$, or active with probability $\phi$. If it is
active, then it can either be damaged, with probability
$1-g_0\left(\Phi \right)$, or usable and belong to finite
usable components with probability $H^{U}_0(1)$. Summing all these
contributions we get
\begin{equation*}
 1=U^{\infty}+1-\phi+\phi \left[1-g_0\left(\Phi \right)+H^{U}_0(1) \right],
\end{equation*}
from which using Eq.~\eqref{eq:H_0} and $H^{U}_1(1)=1-u$ we recover Eq.~(2) in the main.

\subsection{Finite damaged components}
We turn now our attention to finite damaged components. We denote by $\mathcal{N}^{D}_s$ the number of finite damaged components, $n_s^D=\mathcal{N}_s^D/(\phi N)$ the density of damaged components per active node. $\pi_s^D=sn_s^D$ is the probability that the size of a finite damaged component to which a randomly chosen active node belongs is $s$. Following the argument of the previous section, we derive recursive equations for the generating function of $\pi_s^D$.

First, we write down an expression for $\pi_s^D$, the
probability that a randomly chosen active node belongs to a finite
damaged component of size $s$. Denoting by $\rho_{s}^D$ the
probability that following a randomly chosen edge leading to an active
node we reach a finite damaged component of size $s$, we can write
\begin{widetext}
\begin{equation}
 \pi_{s}^D = \sum_{k}p_k \sum_{s_1,\dots,s_k} \delta_{s_1+\dots+s_k,s-1}\sum_{n=0}^{k-1}{k \choose n}\phi^n(1-\phi)^{k-n}[1-(1-\psi)^{k-n}] \prod_{i=1}^n\left(\rho_{s_i}^D \right)\prod_{j=n+1}^{k} \left(\delta_{s_j,0}\right),
\end{equation}
\end{widetext}
where the sum over the number of active neighbors runs from $n=0$ to
$k-1$, since an active node surrounded by active neighbors is
usable. However, since the term with $k=n$ vanishes, we can extend the
sum up to $k$. The term $[1-(1-\psi)^{k-n}]$ is the probability that
at least one of the $k-n$ inactive neighbors corrupts the focal
node. Following the same computation of the previous section, we
multiply by $z^s$ and sum over $s$ to get
\begin{equation}
H_0^D(z)  = z \left[g_0\left(1-\phi + \phi H_{1}^D(z)\right) - g_0\left(\Phi- \phi +\phi H_{1}^D(z)\right) \right].
\label{eq:recursive_small_unusable_0}
\end{equation}

Following a link to an active node, we can either end up in an usable
node, hence reaching a finite usable component of size $0$, with
probability $\rho_0^D = g_1(\Phi)$, or lead to a finite damaged
component of size $s>0$ with probability
\begin{widetext}
\begin{equation}
\rho_s^D = \sum_{r}q_{r} \sum_{s_{1},\cdots,s_{r}} \delta_{s_{1}+\cdots+s_{r},s-1}\sum_{n=0}^{r-1}{r \choose n}\phi^n(1-\phi)^{r-n}[1-(1-\psi)^{r-n}] \prod_{i=1}^{n}{\left( \rho_{s_{i}}^D\right)}\prod_{j=n+1}^{r} \left(\delta_{s_{j},0}\right).
\end{equation}
\end{widetext}
We can extend the sum over $n$ up to $r$ also in this case.
Multiplying by $z^s$ and summing over $s$ we get
\begin{equation}
 \label{eq:recursive_small_unusable_1}
H_1^D(z)  = g_1(\Phi)
 + z \left[g_1\left(1-\phi+\phi H_{1}^D(z)\right) - g_1\left(\Phi- \phi + \phi H_{1}^D(z)\right) \right].
\end{equation}
Again, a consistency check using Eq.~\eqref{eq:recursive_small_unusable_0} and $H_1(1)=1-d$ allows us to recover Eq.~(4) in the main.

\subsection{Average size of finite components}

Now we can compute the average size of finite usable clusters
\begin{equation}
 \langle s \rangle_{U} = \frac{\sum_{s}s \pi_s^{U}}{\sum_{s}\pi^{U}_s} = \frac{\partial_z H_0^{U}(1)}{H_0^U(1)},
\end{equation}
which in continuous percolation transitions behaves, close to the percolation threshold $\phi_c$, as
$\langle s\rangle -1 \sim |\phi - \phi_c|^{-\gamma}$.
Taking a derivative of Eq.~\eqref{eq:H_0} with respect to $z$ and evaluating in $z=1$, using
$g_0'(z)=\langle k \rangle g_1(z)$, we get
\begin{equation*}
 \langle s \rangle _{U} = 1 + \frac{\phi \langle k \rangle g_1\left(\Phi-\phi u  \right) \partial_z H_1^{U}(1)}{H_0^U(1)}.
\end{equation*}
$\partial_z H_1^{U}(1)$ can be found taking a derivative of Eq.~\eqref{eq:H_1}, which gives
\begin{equation*}
 \partial_z H_1^{U}(1)=g_1\left(\Phi -\phi u \right)+\phi g_1'\left(\Phi -\phi u \right) \partial_z H_1^{U}(1).
\end{equation*}
This linear equation for $\partial_z H_1^{U}(1)$ can be easily solved. We finally obtain
\begin{equation}
 \langle s \rangle_{U} = 1 + \frac{\phi \langle k \rangle [g_1\left(\Phi\right)-u]^2}{H_0^U(1)[1-\phi g_1'\left(\Phi - \phi u\right)]}.
 \label{eq:uean_cluster_size}
\end{equation}

\begin{figure}
\center
\includegraphics[width=0.9\textwidth]{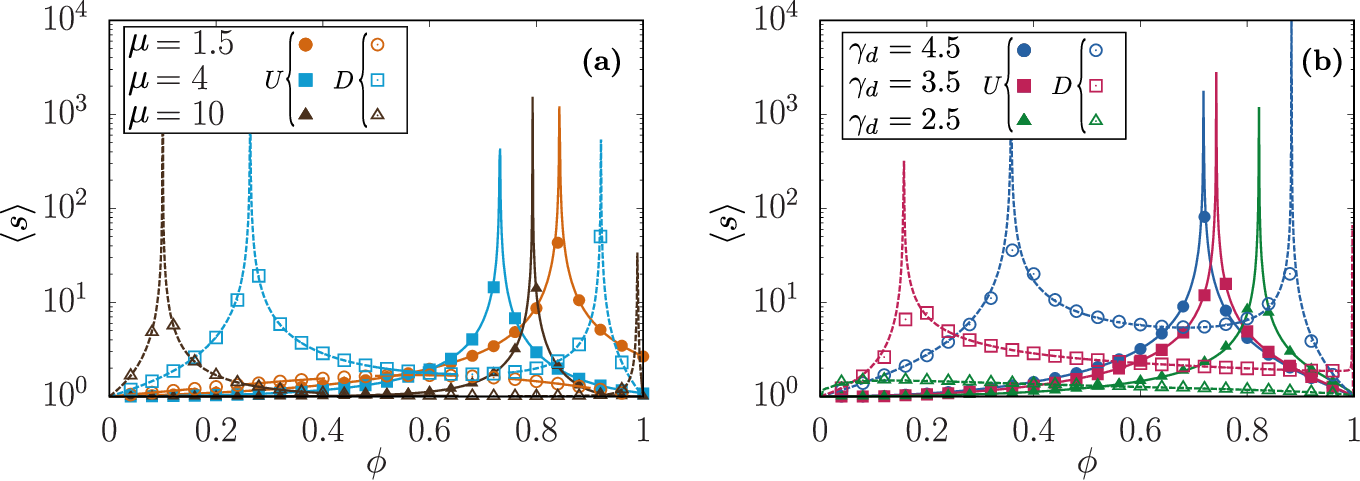}
    \caption{Average usable $\langle s \rangle_{U}$ and damaged $\langle s \rangle_{D}$ cluster sizes for $\psi=1$ from the solution of Eq.~\eqref{eq:uean_cluster_size} (continuous lines) and Eq.~\eqref{eq:dean_cluster_size} (dashed lines), respectively. Symbols are numerical simulations on networks of $N=10^6$ nodes, averaged over $100$ realizations. Filled symbols are for $\langle s \rangle_{U}$, empty symbols are for $\langle s \rangle_{D}$. (a) ER networks with various average degree $\mu$. (b) PL networks with $\km=3$. Compare with Fig.~2 in the main where the order parameters $U^{\infty},D^{\infty}$ are shown for the same networks.}
    \label{fig:susceptibility}
\end{figure}

Concerning finite damaged components, in a perfectly analogous manner we obtain,
for the average finite damaged components size $\langle s \rangle_{\text{D}} = \partial_z H_0^{D}(1)/H_0^D(1)$
\begin{equation}
 \langle s \rangle_{D} = 1 + \frac{\phi \langle k \rangle [1-g_1(\Phi)-d]^2}{H_0^D(1)\left\{1- \phi[g_1'(1-\phi d)-g_1'(\Phi-\phi d)] \right\}}.
 \label{eq:dean_cluster_size}
\end{equation}
See Fig.~\ref{fig:susceptibility} for the numerical solution of Eq.~\eqref{eq:uean_cluster_size} and Eq.~\eqref{eq:dean_cluster_size} with a comparison with numerical simulations. The presence of two peaks in $\langle s \rangle_{D}$ marks the reentrant damaged percolation transition.

\section{Additional details on the usable and damaged transitions}

\subsection{The phase transition of the usable component}
\label{sec:critical_properties_usable}
We report here for better clarity the equation for the order parameter $U^{\infty}$ (see Eq.~(2) in the main)
\begin{equation}
 U^{\infty} = \phi \left[g_0\big(\Phi \big)-g_0\big(\Phi - \phi u \big) \right],
 \label{eq:U_infty}
\end{equation}
and the recursive equation for $u$ (see Eq.~(3) in the main)
\begin{equation}
u =  g_1 \big(\Phi \big) - g_1 \big(\Phi - \phi u \big).
\label{eq:u_psi}
\end{equation}
As we have already discussed, $u=0$ is always a solution. To find
another solution $u>0$, we can assume $u \ll 1$ and expand the
r.h.s. of Eq.~\eqref{eq:u_psi} to get at leading order
\begin{equation}
\label{eq:expansion_usable}
 \phi^2g_1 ''(\Phi) u^2 - 2[\phi g_1'(\Phi)-1]u \simeq 0,
\end{equation}
from which it follows
\begin{align}
 u \simeq \begin{cases}
  0 \quad &\phi g_1'(\Phi)-1 \leq 0,\\
  \frac{2[\phi g_1'(\Phi) - 1]}{\phi^2g_1''(\Phi)} \quad &\phi g_1'(\Phi) - 1 > 0.
 \end{cases}
 \label{eq:u_close}
\end{align}
This shows that a continuous transition takes place at the critical
threshold $\phi_c^U$, solution of the equation (see Eq.~(6) in the main)
\begin{equation}
 1=\phi_c^U g_1'(\Phi_c^U).
 \label{eq:critical_line_U}
\end{equation}
where $\Phi_c^U=1-\psi(1-\phi_c^U)$, as we have already noted in the main using a stability argument.

\subsubsection{Usable transition in the limit $\psi \to 0$}
For $\psi=0$, NIDP reduces to standard site percolation, and
Eq.~\eqref{eq:critical_line_U} should reduce to the well-known
percolation threshold $\phi_c^{\text{site}}=1/b$, where
$b=\langle k(k-1)\rangle/\langle k \rangle$.
The degree heterogeneity comes
then into play when taking the limit $\psi \to 0$. For homogeneous
networks and power-law networks with $\gamma_d>4$, assuming $\psi \ll
1$ we have from Eq.~\eqref{eq:critical_line_U}
\begin{equation}
\phi_c^U \simeq \phi_c^{\text{site}} + \mu_3 \left(\phi_c^{\text{site}}\right)^2 (1-\phi_c^{\text{site}})\psi,
\label{eq:threshold_homogeneous}
\end{equation}
where $\mu_3=\langle k(k-1)(k-2) \rangle$. For heterogeneous networks
with $2 < \gamma_d <4$ the divergence of $\langle k(k-1)(k-2) \rangle$,
for $3< \gamma_d <4$, and also of $ \langle k(k-1) \rangle$, for
$2<\gamma_d <3$, makes this limit less trivial. In fact, when $\psi
\to 0$, $\Phi=1-(1-\phi)\psi \to 1$ and the singular behavior of
$g_1'(1-x)$ for $x\to 0^{+}$ must be taken into account. In
particular, we can write
\begin{align}
 \nonumber
 g_1'\left(1-(1-\phi_c^U)\psi \right) \simeq [b - (\gamma_d-2)c_2[(1-\phi_c^U)\psi]^{\gamma_d-3}],
\end{align}
where $c_2$ is a constant which depends on the details of the degree
distribution, see~\cite{cirigliano2023extended}. Using this expansion,
from Eq.~\eqref{eq:critical_line_U} it follows, at leading order in
$\psi$,
\begin{align}
\phi_c^U \simeq \begin{cases}
               \phi_c^{\text{site}}+\kappa_1 \psi^{\gamma_d-3},\quad &3<\gamma_d<4\\
               \kappa_2\psi^{3-\gamma_d}, \quad &2<\gamma_d<3
              \end{cases}
              \label{eq:threshold_PL}
\end{align}
where $\kappa_1=\left(\phi_c^{\text{site}}\right)^2
(1-\phi_c^{\text{site}})(\gamma_d-2)c_2$,
$\kappa_2=-c_2(\gamma_d-2)$.
See the box inside Fig.~4(a) in the main for the thresholds $\phi_c^{U}$ for $\psi \ll 1$.

\subsection{The phase transition of the damaged component}
\label{sec:critical_properties_damaged}
We report here for better clarity the equation for the order parameter $D^{\infty}$ (see Eq.~(4) in the main)
\begin{equation}
 D^{\infty}=\phi \left[1-g_0(\Phi)+g_0(\Phi-\phi d)-g_0(1-\phi d) \right].
 \label{eq:D_infty}
\end{equation}
and the recursive equation for $d$ (see Eq.~(5) in the main)
\begin{equation}
d = 1-g_1(1-\phi d) + g_1(\Phi-\phi d) - g_1(\Phi) .
\label{eq:d_psi}
\end{equation}
As we have already discussed, $d=0$ is always a solution.
To find another solution $d>0$, we expand for
  $d \ll 1$ to get
\begin{equation}
 c_2 \phi^{\gamma_d-2} d^{\gamma_d-3} + \frac{1}{2}\left[\mu_3-g_1''(\Phi) \right]\phi^2 d + \left\{1-\phi [b-g_1'(\Phi)] \right\} \simeq 0,
 \label{eq:d_close}
\end{equation}
where $\mu_3 = \langle k (k-1)(k-2) \rangle/\langle k \rangle$, and $c_2$ is a constant depending on $\gamma_d$ and in general on the small-degree part of the degree distribution.
In contrast to the usable transition, here the
asymptotic expansions of generating functions of power-law degree
distributions close to the singularity in $1$ are used. This leads to
$\gamma_d$-dependent behavior of the solution.

For $\gamma_d>4$, as for homogeneous degree distributions,
we can ignore the first, subleading, term on
the l.h.s. of Eq.~\eqref{eq:d_close} to get
\begin{align}
 d \simeq \begin{cases}
  0 \quad & \phi [b-g_1'(\Phi)]-1 \leq 0,\\
  \frac{2\{\phi[b-g_1'(\Phi)]-1\}}{\phi^2[\mu_3-g_1''(\Phi)]} \quad & \phi [b-g_1'(\Phi)] -1 > 0.
 \end{cases}
 \label{eq:d_homogeneous}
\end{align}

For $3<\gamma_d<4$ instead, the second-order term is subleading, and we must keep the first and the last
 terms on the l.h.s. of Eq.~\eqref{eq:d_close}, from which we get
\begin{align}
 d \simeq \begin{cases}
  0 \quad & \phi [b-g_1'(\Phi)]-1\leq 0,\\
  \left[\frac{\{\phi[b-g_1'(\Phi)]-1\}}{c_2\phi^{\gamma_d-2}}\right]^{1/(\gamma_d-3)} \quad & \phi [b-g_1'(\Phi)] - 1  > 0.
 \end{cases}
 \label{eq:d_weakly_het}
\end{align}

Finally, for $2<\gamma_d<3$ the branching factor diverges, $b$ is a
negative coefficient, coming from the asymptotic expansion of the
generating functions, and $c_2<0$.
Hence, as long as $0< \phi<1$, we can write
\begin{equation}
d \simeq \left[\frac{\{1-\phi[b-g_1'(\Phi)]\}}{-c_2\phi^{\gamma_d-2}}\right]^{1/(\gamma_d-3)},
\label{eq:d_strongly_het}
\end{equation}
since the condition $\phi[b-g_1'(\Phi)] - 1 \leq 0$ is always satisfied.
Some care must be taken for $\phi \to 0^{+}$ and for $\phi \to 1^{-}$, in
the case of strongly heterogeneous networks. We discuss this point below.

\subsection{Conditions for the existence of a GDC}

As we have discussed in the previous subsection and in Sec.~II.B in the main, for $\gamma_d>3$ a continuous transition for the size of the GDC takes place at the percolation thresholds determined by the solution of the equation
\begin{equation}
 1=\phi_{\pm}^{D}\left[b-g_1'(\Phi_{\pm}^{D})\right].
 \label{eq:critical_line_D}
\end{equation}
The existence of two distinct solutions, i.e., the existence of a finite interval $(\phi_{-}^D,\phi_{+}^D)$ in which $D^{\infty}>0$, crucially depends on the values of the networks parameters $\underline{\lambda}$. Denoting with $\underline{\lambda}_{*}$ the values of the parameters for which Eq.~\eqref{eq:critical_line_D} admits two coincident solutions $\phi_*=\phi_{-}^D=\phi_{+}^D$, we can write two conditions for $\phi_*$ and $\underline{\lambda}_{*}$
\begin{align}
\label{eq:curve_1}
f(\phi_*, \underline{\lambda}_{*}) &= \phi_{*} g_{1}'(\Phi_{*},\underline{\lambda}_{*}) - \phi_{*}b(\underline{\lambda}_{*}) + 1 = 0,\\
\label{eq:curve_2}
\partial_{\phi}f(\phi_*, \underline{\lambda}_{*}) &= g_{1}'(\Phi_{*},\underline{\lambda}_{*}) + \psi \phi_{*} g_{1}''(\Phi_{*},\underline{\lambda}_{*}) - b(\underline{\lambda}_{*}) = 0.
\end{align}
Thus $\underline{\lambda}_{*}$ is the extremal choice of the network parameters for the existence of a GDC. Below, we provide the explicit expressions for Eqs.~\eqref{eq:curve_1},\eqref{eq:curve_2} for the network classes of RRN, ER, and PL considered in the main.

\subsubsection{Random regular networks}
For RRN with degree $\mu$, hence $g_0(z)=z^{\mu}$ and $b=\mu-1$, the two equations \eqref{eq:curve_1},\eqref{eq:curve_2} read
\begin{align}
\label{eq:mu_star_RRN1}
 \phi_{*} [1-(1-\phi_*)\psi]^{\mu_{*}-2} - \phi_*(\mu_{*}-1) + 1  &=0,\\
 \label{eq:mu_star_RRN2}
 [1-(1-\phi_*)\psi]^{\mu_{*}-2}+ \psi \phi_{*} (\mu_{*}-2)[1-(1-\phi_*)\psi]^{\mu_{*}-3}  - (\mu_{*}-1) &=0.
\end{align}
These equations can be solved numerically. Note that for a nontrivial solution to exist we need $\mu_{*}>3$, otherwise the equations cannot be both satisfied. Hence we can conclude that $\mu_{*} \geq 4$. In particular, for $\psi=1$ we have $\mu_{*}=4$, and decreasing $\psi$ the value of $\mu_{*}$ increases, as shown in Fig.~4(b) in the main.

\subsubsection{Erd\H{o}s-R\'enyi networks}
For ER networks with average degree $\mu$ we can solve these nonlinear equations explicitly. Using $g_0(z)=e^{\mu(z-1)}$ and $b=\mu$ we have
\begin{align}
\label{eq:mu_star_ER1}
\phi_{*}\mu_{*}e^{\mu_{*}\psi(\phi_*-1)}-\phi_{*} \mu_{*}+1&=0,\\
\label{eq:mu_star_ER2}
e^{\mu_{*}\psi(\phi_*-1)}+\psi \mu_{*}\phi_* e^{\mu_{*}\psi(\phi_*-1)}-1&=0.
\end{align}
From the second equation we get
\begin{equation}
e^{\psi \mu_{*}(\phi_{*}-1)} = \frac{1}{1+\psi \phi_{*}\mu_{*}},
\label{eq:passaggio}
\end{equation}
which plugged into the first one gives, setting $x=\phi_{*}\mu_{*}$,
\[x^2 - x - 1/\psi = 0. \]
The positive solution (since $x=\phi_{*}\mu_{*}$ must be positive) of this equation is
\begin{equation}
 \phi_{*}\mu_{*}=x=\varphi(\psi)=\frac{1+\sqrt{1+4/\psi}}{2}.
\end{equation}
We use the symbol $\varphi(\psi)$ to denote this solution since $\varphi(1)=(1+\sqrt{5})/2$ is the golden ratio, usually denoted by $\varphi$. Substituting this value in Eq.~\eqref{eq:passaggio}, we get $1+\psi \varphi(\psi)=e^{\psi \mu_{*}}e^{-\psi \varphi(\psi)}$, from which we can solve for $\mu_{*}$ to get (see Eq.~(14) in the main)
\begin{equation}
\mu_{*} = \varphi(\psi) + \frac{\ln \left[1+\psi \varphi(\psi)\right]}{\psi},
\label{eq:mu_star_ER}
\end{equation}
which is shown in Fig.~4(b) in the main. The inverse function of Eq.~\eqref{eq:mu_star_ER}, $\psi_{*}(\mu)$, gives the minimum value of $\psi$ needed at fixed $\mu$ to observe a reentrant phase. The activation probability at which the two solutions appear is given by
\begin{equation}
 \phi_{*} = \frac{\psi \varphi(\psi)}{\varphi(\psi)+\ln \left[1+\psi \varphi(\psi) \right]}.
\end{equation}

\subsubsection{Power-law degree distributions}
For power-law degree distributions $p_k = k^{-\gamma_d}/\zeta(\gamma_d,\km)$, we can write the generating function $g_0(z)$ in terms of the Lerch transcendent, a special function defined by\footnote{The Lerch trascendent is typically denoted by $\Phi$. Here we use the symbol $\mathcal{L}$ to avoid ambiguities with $\Phi=1-(1-\phi)\psi$.}
\begin{equation}
 \mathcal{L}(z,s,\alpha) = \sum_{n=0}^{\infty} \frac{z^s}{(n+\alpha)^{s}}.
\end{equation}
We have
\begin{equation}
 g_0(z)=z^{\km}\frac{\mathcal{L}(z,\gamma_d, \km)}{\zeta(\gamma_d, \km)}.
\end{equation}
Knowing the derivatives of the Lerch transcendent, we can compute the derivative of $g_0(z)$. In particular, we can use the formula
\begin{equation}
 \frac{d}{dz}\left[z^{a}\mathcal{L}(z,s,b)\right] = (a-b)z^{a-1}\mathcal{L}(z,s,b)+z^{a-1}\mathcal{L}(z,s-1,b),
\end{equation}
from which we get
\begin{align}
 g_1(z)&=z^{\km-1}\frac{\mathcal{L}(z,\gamma_d-1, \km)}{\zeta(\gamma_d-1, \km)},\\
 g_1'(z)&=\frac{z^{\km-2}\mathcal{L}(z,\gamma_d-2,\km)-z^{\km-2}\mathcal{L}(z,\gamma_d-1,\km)}{\zeta(\gamma_d-1,\km)},\\
 g_1''(z)&=\frac{z^{\km-3}\mathcal{L}(z,\gamma_d-3,\km)-3z^{\km-3}\mathcal{L}(z,\gamma_d-2,\km) +2z^{\km-3}\mathcal{L}(z,\gamma_d-1,\km)}{\zeta(\gamma_d-1,\km)}.
\end{align}
An explicit solution of Eqs.~\eqref{eq:curve_1},\eqref{eq:curve_2} is clearly not available in this case. However, we can solve these equations numerically to find $\gamma_{d*}$ for various $\km$. Results are shown in Fig.~4(c),(d) in the main.

\section{Critical exponents}

\subsection{Exponents of the usable transition}

The expansion Eq.~\eqref{eq:u_close} derived above is valid for any arbitrary degree distribution as soon as $\psi>0$ and for $\phi<1$. For $\psi=0$ we recover the expansions in standard site percolation~\cite{cirigliano2024scaling}. Hence for any fixed $\psi>0$, the curve $\phi_c^{U}(\psi)$, solution of Eq.~\eqref{eq:critical_line_U}, determines a critical line in the
$\phi-\psi$ plane. Setting $t=\phi-\phi_c^U$ and $\Phi_c^U=1-(1-\phi_c^U)\psi$, we can expand Eq.~\eqref{eq:u_psi} for $0<t \ll 1$ to get
\begin{align}
u \simeq \frac{2[g_1'(\Phi_c^U)+\psi \phi_c^U  g_1''(\Phi_c^U)]}{(\phi_c^U)^2 g_1''(\Phi_c^U)}t.
\label{eq:asymptotics_u_psi}
\end{align}
For $0<t\ll 1$, inserting Eq.~\eqref{eq:asymptotics_u_psi} into Eq.~\eqref{eq:U_infty}, at lowest order we get
\begin{equation}
 U^{\infty} \simeq  \frac{2 (\phi_c^{U})^2 \langle k  \rangle [g_1'(\Phi_c^U)+\psi \phi_c^U g_1''(\Phi_c^U)]}{g_1''(\Phi_c^U)}t,
\end{equation}
from which we conclude that $U^{\infty}/\phi \sim t^{\beta^U}$ with
$\beta^U=1$, the homogeneous mean-field percolation exponent~\cite{cirigliano2024scaling}.
Note that this result is valid also for heterogeneous networks
with a power-law degree distribution $p_k \sim k^{-\gamma_d}$, in contrast with the
$\gamma_d$-dependent exponent found for standard percolation on
heterogeneous networks with $2<\gamma_d<4$~\cite{cirigliano2024scaling}.

From the expression for $\langle s \rangle_{U}$ in Eq.~\eqref{eq:uean_cluster_size} we can get the critical exponent $\gamma$. Approaching $\phi_c^{U}$, the term in the square brackets in the denominator vanishes and then $\langle s \rangle_{U}$ diverges. Expanding the denominator for small $t=\phi - \phi_c^{U}$ we get at lowest order, using the condition for the threshold $1=\phi_c^{U}g_1'(\Phi_c^{U})$,
\begin{equation}
1-\phi g_1'(\Phi - \phi m) \simeq - t [g_1'(\Phi^{U}_c)+\psi \phi_c^{U} g_1''(\Phi^{U}_c)]+(\phi_c^{U})^2g_{1}''(\Phi_c^{U}) u.
\end{equation}
For $t<0$ we have $u=0$, from which it follows $\langle s \rangle \sim (-t)^{-1}$, hence $\gamma=1$. For $t>0$, we can use Eq.~\eqref{eq:asymptotics_u_psi} to get
\begin{equation}
 - t [g_1'(\Phi^{U}_c)+\psi \phi_c^{U} g_1''(\Phi^{U}_c)]+(\phi_c^{U})^2g_{1}''(\Phi_c^{U}) u  \simeq t [g_1'(\Phi^{U}_c)+\psi \phi_c^{U} g_1''(\Phi^{U}_c)] \sim t,
\end{equation}
from which we can conclude that $\gamma=1$. Again, the critical exponent $\gamma$ does not depend on the degree distribution even for scale-free networks.

We can conclude that the usable percolation transition on random graphs of arbitrary degree distribution belongs, even for PL with $2<\gamma_d<4$, to the universality class of percolation in homogeneous networks.

\subsection{Exponents of the damaged transition}

While we have found a universal behavior of the usable transition independently of the degree distribution, for the damaged phase transition the critical exponents may depend on $\gamma_d$, as it happens for standard percolation. We have learnt that we have at most two distinct percolation thresholds $\phi_{\pm}^{D}$ corresponding to the birth and the dismantling of a GDC. The expansions must then be carried out for $t_{-}=\phi-\phi_{-}^{D} \ll 1$ and for $t_{+}=\phi_{+}^{D}-\phi \ll 1$. The behavior of the order parameter $D^{\infty}$ close to the threshold can be obtained by expanding Eq.~\eqref{eq:D_infty} for small $d$ to get, using $g_1(z)=g_0'(z)/\langle k \rangle$,
\begin{equation}
 D^{\infty} \simeq \phi^2 \langle k \rangle \left[1-g_1(\Phi) \right]d.
 \label{eq:D_close_to_threshold}
\end{equation}
Note that Eq.~\eqref{eq:D_close_to_threshold} is valid only if $\phi<1$ and $\psi>0$. Additional care must be taken, as we detail below, when dealing with the case $\phi \to 1$.
For homogeneous degree distributions and power-laws with $\gamma_d>4$, we can expand Eq.~\eqref{eq:d_homogeneous} to get
\begin{equation}
d \simeq \frac{2 [b-g_1'(\Phi_{\pm}^{D})-\psi \phi_{\pm}^{D} g_1''(\Phi_{\pm}^{D})]}{\left(\phi_{\pm}^{D}\right)^2}t_{\pm} \sim t_{\pm},
\end{equation}
from which we can conclude that both the exponents $\beta^{D}_{-}$ and $\beta^{D}_{+}$ describing the behavior of $D^{\infty}/\phi$ close to $\phi_{-}^{D}$ and $\phi_{+}^{D}$, respectively, take the mean-field value $\beta^{D}_{-}=\beta^{D}_{+} = 1$.

For power-law degree distributions with $3<\gamma_d<4$, we can expand Eq.~\eqref{eq:d_weakly_het} to get
\begin{equation}
 d \simeq \left[ \frac{[b-g_1'(\Phi_{\pm}^{D})-\psi \phi_{\pm}^{D} g_1''(\Phi_{\pm}^{D})]}{ c_2\left( \phi_{\pm}^{D}\right)^{\gamma_d-2}}t_{\pm} \right]^{1/(\gamma_d-3)} \sim t_{\pm}^{1/(\gamma_d-3)},
\end{equation}
from which we get the percolation critical exponents $\beta^{D}_{-}=\beta^{D}_{+}=1/(\gamma_d-3)$, the same values for standard site percolation. This happens because hubs typically belong to the the GDC. As we have seen in the previous section, the thresholds $\phi_{-}$ and $\phi_{+}$ tend to $0$ and $1$, respectively, as soon as $\gamma_d \to 3^{+}$. However, since they are still in the range $0<\phi_{\pm}^{D}<1$, the expansions above can be carried out avoiding the singularity in $z=1$ of the generating function $g_1'(z)$, since $\Phi_{-}^{D}=1-(1-\phi_{-}^{D})\psi < 1$ and $\Phi_{+}^{D} = 1-(1-\phi_{+}^{D})\psi <1$.

For power-law degree distributions with $2<\gamma_d<3$ instead, Eq.~\eqref{eq:critical_line_D} does not admit any solution, and the birth and dismantling of the GDC take place for $\phi=0$ and $\phi=1$, respectively. This is similar to what happens in standard percolation on scale-free networks for $\phi \to 0$. To get the behaviour of $d$ close to these thresholds we must proceed carefully. For $\phi$ close to $0$, Eq.~\eqref{eq:d_strongly_het} is still valid and we can expand its r.h.s. for $t_{-}=\phi \ll 1$. We get
\begin{equation}
 d \simeq  (-c_2)^{\frac{1}{3-\gamma_d}} t_{-}^{\frac{\gamma_d-2}{3-\gamma_d}}.
\end{equation}
From Eq.~\eqref{eq:D_close_to_threshold} we then get, recalling that $\phi=t_{-}$
\[D^{\infty}/\phi \simeq t_{-} \langle k \rangle [1-g_1(1-\psi)] (-c_2)^{\frac{1}{3-\gamma_d}} t_{-}^{\frac{\gamma_d-2}{3-\gamma_d}} \sim t_{-}^{1/(3-\gamma_d)},\]
from which we get $\beta^{D}_{-}=1/(3-\gamma_d)$. For $\phi \to 1^{-}$ instead we cannot use Eq.~\eqref{eq:d_strongly_het}, since it has been obtained for $0<\phi<1$ and $d \ll 1$, and now we are expanding for both $t_{+}=1-\phi \ll 1$ and $d \ll 1$. Setting $\Phi=1-\psi t_{+}$ and using the asymptotic expansion of $g_1$ from Eq.~\eqref{eq:d_psi} we get
\begin{equation}
 d \simeq -c_2 \phi^{\gamma_d-2} d^{\gamma_d-2} + c_2\left(\psi t_{+} + \phi d \right)^{\gamma_d-2} - c_2 \psi^{\gamma_d-2}t_{+}^{\gamma_d-2}.
\end{equation}
The only meaningful solution of this equation can be derived assuming $\psi t_{+} \ll \phi d$ and using the binomial expansion, at leading order
\[d \simeq c_2 (\gamma_d-2) \phi^{\gamma_d-3} d^{\gamma_d-3} \psi t_{+} -c_2 \psi^{\gamma_d-2} t^{\gamma_d-2}, \]
from which we finally get
\begin{equation}
 d \simeq -c_2 \psi^{\gamma_d-2}t^{\gamma_d-2}_{+}.
 \label{eq:d_phi_to_1}
\end{equation}
As noted above, we cannot use Eq.~\eqref{eq:D_close_to_threshold} since now $d \ll1 $ but also $\phi \to 1$. Starting from Eq.~\eqref{eq:D_infty}, we can expand its r.h.s. for $d \ll 1$ and $t_{+}\ll 1 $ to get
\begin{align*}
 D^{\infty}/\phi \simeq 1-\left(1-\langle k \rangle \psi t_{+} +c_1 \psi^{\gamma_d-1} t_{+}^{\gamma_d-1}\right) + 1-\langle k \rangle \left(\psi t_{+} + \phi d \right) + c_1 \left(\psi t_{+} + \phi d \right)^{\gamma_d-1} - \left(1- \langle k \rangle \phi d + c_1 \psi^{\gamma_d-1} t_{+}^{\gamma_d-1} \right).
\end{align*}
We see from this expression that the constants and the linear terms in $t_{+}$ and in $d$ cancel out. Using $t_{+} \ll \phi d $ and Eq.~\eqref{eq:d_phi_to_1}, we can further expand to finally get
\begin{equation}
 D^{\infty} / \phi \simeq (\gamma_d-1)c_1 \phi^{\gamma_d-2} d^{\gamma_d-2} \psi t_{+} \simeq (\gamma_d-1)c_1 \left[(-c_2)^{\gamma_d-2} \psi^{1+(\gamma_d-2)^2} \right] t_{+}^{1+(\gamma_d-2)^2},
\end{equation}
from which we get $\beta_{+}^{D}=1+(\gamma_d-2)^2$.

We turn now our attention to the critical exponent $\gamma^{D}$. We begin considering the case in which $0<\phi_{-}^{D}\leq \phi_{+}^{D}<1$, that is for homogeneous degree distributions and power-laws with $\gamma_d>3$. From Eq.~\eqref{eq:dean_cluster_size}, we see a divergence approaching $\phi_{-}^{D}$ and $\phi_{+}^{D}$, as the denominator vanishes. We now carefully expand the denominator for $t_{-}=\phi-\phi_{-}^{D}$ and $t_{+}=\phi_{+}^{D}-\phi$. Note that we have chosen the signs in order to have both $t_{-}>0$ and $t_{+}>0$ in the phase with a GDC, hence when $d>0$.

We first consider homogeneous degree distributions or power-laws with $\gamma_d>4$. We have
\begin{align*}
 1-\phi \left[g_1'(1-\phi d) - g_1'(\Phi-\phi d) \right] &\simeq 1-\phi \left[ b - \mu_3 \phi d - g_1'(\Phi) + g_1''(\Phi)\phi d \right] = 1-\phi \left[b - g_1'(\Phi) \right] + \phi^2 d [\mu_3-g_1''(\Phi)]\\
 &\simeq \left \{ \phi \left[b - g_1'(\Phi) \right]-1 \right\} \left[2 \theta(\phi \left[b - g_1'(\Phi) \right]-1 ) - 1 \right],
\end{align*}
where the Heaviside function ($\theta(x) = 1$ if $x>0$ and zero otherwise) allows us to distinguish between the cases with or withouth a GDC. For $t_{-}<0$ and $t_{+}<0$, no GDC is present. Expanding the r.h.s. we get
\[-\left \{ \phi \left[b - g_1'(\Phi) \right]-1 \right\} \simeq -\left[b-g_1'(\Phi_{\pm}^{D}) - \psi \phi_{\pm}^{D}g_1''(\Phi_{\pm}^{D}) \right] t_{\pm}\ \sim -t_{\pm},\]
implying that $\gamma_{\pm}=1$ in the phase with no GDC. For $t_{-}>0$ and $t_{+}>0$, we get
\[\left \{ \phi \left[b - g_1'(\Phi) \right]-1 \right\} \simeq  \left[b-g_1'(\Phi_{\pm}^{D}) - \psi \phi_{\pm}^{D}g_1''(\Phi_{\pm}^{D}) \right] t_{\pm}\ \sim t_{\pm},\]
from which we conclude that $\gamma_{\pm}=1$.

For $3<\gamma_d<4$ care must be taken, since the expansion of $g_1'$ must consider a divergence of the third moment. We can expand the denominator in Eq.~\eqref{eq:dean_cluster_size} to get
\begin{align*}
 1-\phi \left[g_1'(1-\phi d) - g_1'(\Phi-\phi d) \right] &\simeq 1-\phi \left[ b - c_2(\gamma_d-2) \phi^{\gamma_d-3} d^{\gamma_d-3} - g_1'(\Phi) + g_1''(\Phi)\phi d \right]\\
 &\simeq 1-\phi \left[ b -  g_1'(\Phi)\right] + c_2(\gamma_d-2) \phi^{\gamma_d-2} d^{\gamma_d-3} \\
 & \simeq \left \{ \phi \left[b - g_1'(\Phi) \right]-1 \right\} \left[(\gamma_d-2) \theta(\phi \left[b - g_1'(\Phi) \right]-1 ) - 1 \right].
\end{align*}
Expanding again for $\phi$ close to $\phi_{\pm}^{D}$, both for the phase with $d=0$ and for the phase $d>0$, we get $\gamma_{\pm}=1$.

For $2<\gamma_d<3$ instead, additional care must be taken since, as already noted above, no phase transition occurs for $0< \phi<1$, and the birth and the dismantling of the GDC take place at $\phi=0$ and $\phi=1$, respectively. We begin considering the behavior of $\langle s \rangle_{D}$ for $t_{-}=\phi \ll 1$. In this case, no divergence is present in the denominator, which goes to a constant, and from the numerator in Eq.~\eqref{eq:dean_cluster_size} we get $\gamma_{-}=-1$, as it happens for standard site percolation in scale-free networks. For $\phi \to 1$, we must carefully expand the numerator and the factor $H_{0}^{D}(1)$ in the denominator in the r.h.s. of Eq.~\eqref{eq:dean_cluster_size}, since they both vanish, while the other term in the denominator goes to $1$ for $t_{+}\to 0$. We can expand using $t_{+}=1-\phi \ll 1 $ to get
\[1-g_1(\Phi)-d = g_1(1-\phi d) - g_{1}(\Phi-\phi d) \simeq 1+c_2 \phi^{\gamma_d-2} d^{\gamma_d-2} - 1 -c_2\left(\psi t_{+} + \phi d \right)^{\gamma_d-2} \simeq -c_2 d^{\gamma_d-3} \psi t_{+}  \]
and
\[H_0^{D}(1)=g_0(1-\phi d) - g_0(\Phi-\phi d) \simeq 1-\langle k \rangle \phi d - 1 + \langle k \rangle (\psi t_{+}+\phi d) = \langle k \rangle \psi t_{+}. \]
From Eq.~\eqref{eq:d_phi_to_1} we finally get
\begin{equation}
 \langle s \rangle_{D} -1 \simeq \frac{(-c_2)^{2(\gamma_d-2)}\psi^{2(\gamma_d-2)^2-1}}{\langle k \rangle} t_{+}^{2(\gamma_d-2)(\gamma_d-3)+1} \sim t_{+}^{-[-1+2(\gamma_d-2)(3-\gamma_d)]},
\end{equation}
from which we get $\gamma^{D}_{+} = -1 + 2(3-\gamma_d)(\gamma_d-2)$. Note that $-1 \leq \gamma^{D}_{+} \leq -1/2$. As expected, no divergence in the average cluster size is observed at $\phi_{+}^{D}=1$.

\begin{figure}
\center
\includegraphics[width=0.95\textwidth]{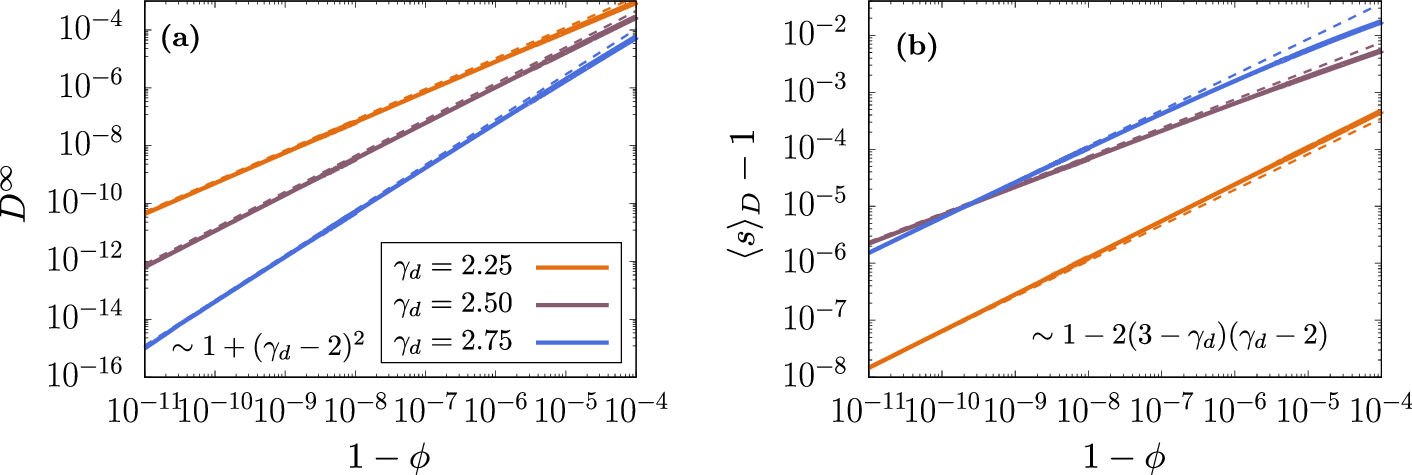}
    \caption{Critical behavior for the damaged percolation transition in scale-free networks with $2<\gamma_d<3$. (a) $D^{\infty}$ from the numerical solution of Eq.~\eqref{eq:D_infty} for $\phi \to \phi_{+}^{D}=1$. The dashed lines are reported for reference and correspond to the scaling with exponents $\beta_{+}^{D}=1+(\gamma_d-2)^2$. (b) $\langle s \rangle_{D}$ from the numerical solution of Eq.~\eqref{eq:dean_cluster_size} for $\phi \to \phi_{+}^{D}=1$. The dashed lines are reported for reference and correspond to the scaling with exponents $-\gamma_{+}^{D}=1-2(3-\gamma_d)(\gamma_d-2)$.}
    \label{fig:strange_exponents_damaged}
\end{figure}

The correctness of the unusual values found for the critical exponents $\beta_+^{D}$ and $\gamma_+^D$ for $2<\gamma<3$ is confirmed in Fig.~\ref{fig:strange_exponents_damaged}, where
analytical predictions are compared to numerical solutions of the equations for $D^{\infty}$
and $\langle s \rangle_D-1$.

\section{Numerical study of NIDP in two-dimensional regular lattices}
\begin{figure}
\center
\includegraphics[width=0.985\textwidth]{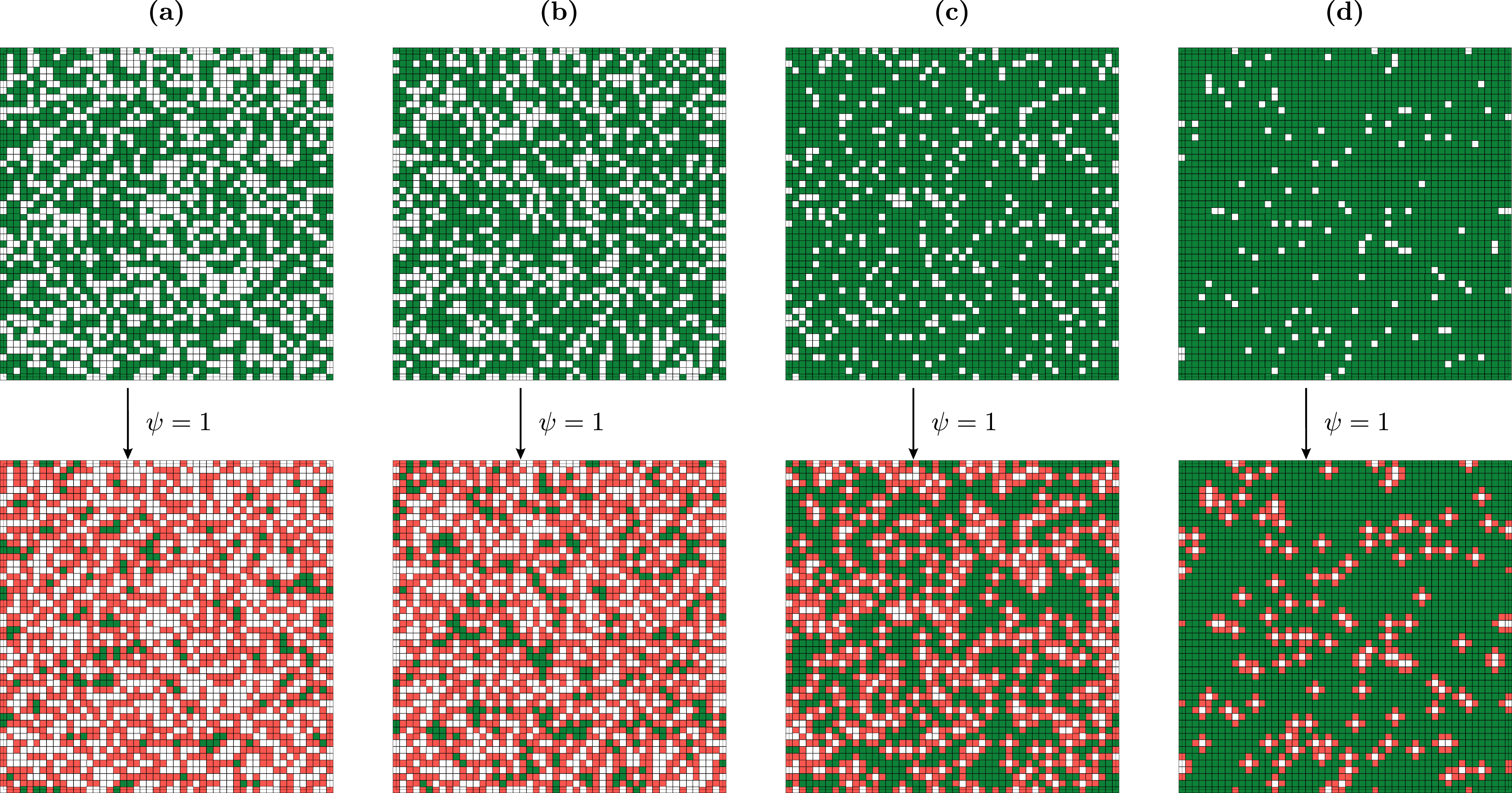}
    \caption{NIDP with $\psi=1$ on a square lattice ($\mu=4$) with $N=2500$ nodes for various values of the activation probability $\phi$. Top: standard site percolation (empty nodes are inactive, green nodes are active). Bottom: the corresponding NIDP with $\psi=1$ (green nodes are usable, red nodes are damaged). (a) $\phi=0.55$, slightly below $\phi_c^{\text{site}}=0.5927\dots$~\cite{christensen2005complexity}. (b) $\phi=0.65$, above the standard percolation threshold. Damaged clusters are large, but no GDC is present. (c) $\phi=0.85$, close to the birth of a GUC. (d) $\phi=0.95$, where a GUC is present. }
    \label{fig:square}
\end{figure}

\begin{figure}
\center
\includegraphics[width=0.985\textwidth]{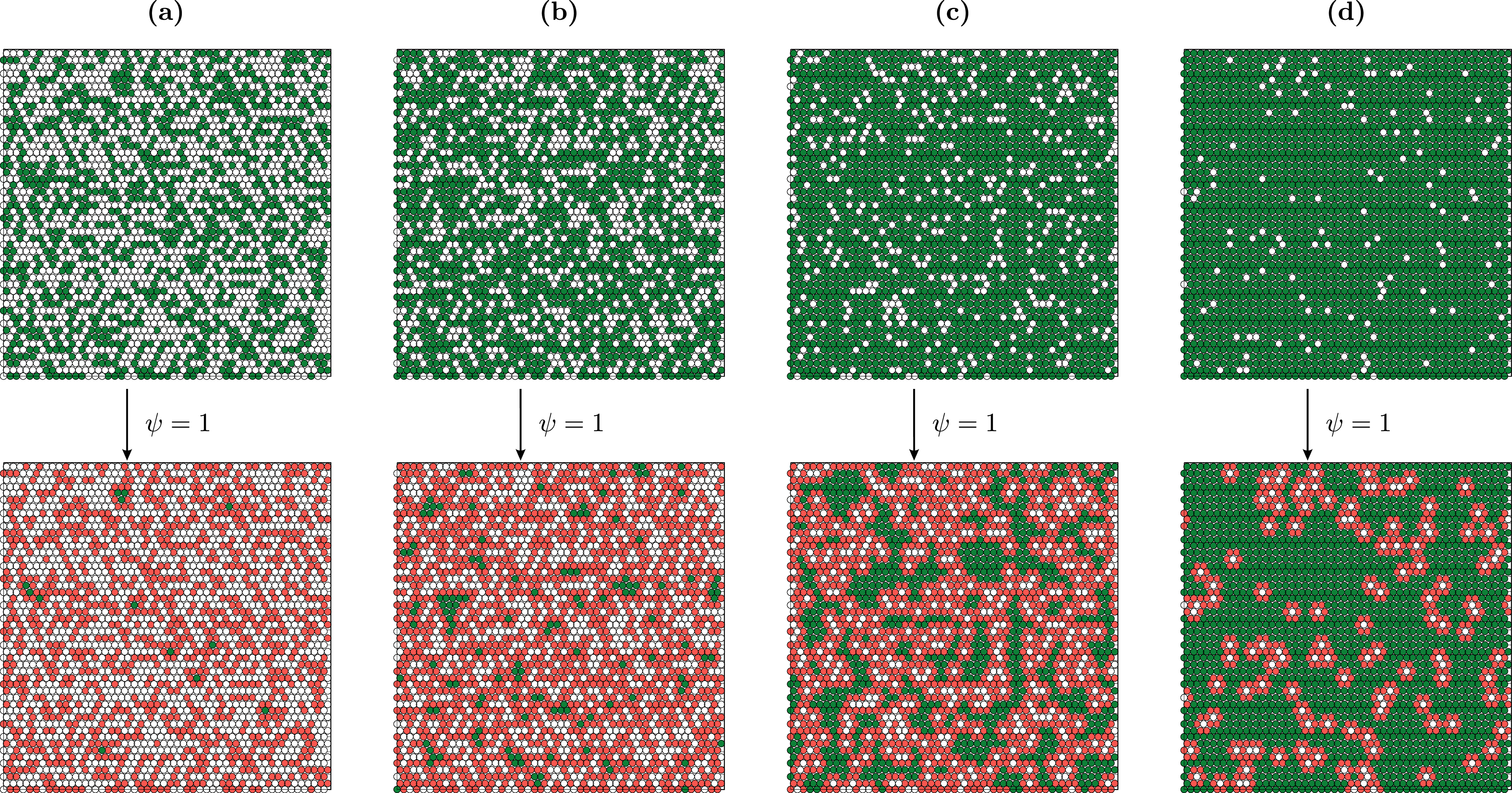}
    \caption{NIDP with $\psi=1$ on a triangular lattice ($\mu=6$) with $N=2500$ nodes for various values of the activation probability $\phi$. Top: standard site percolation (empty nodes are inactive, green nodes are active). Bottom: the corresponding NIDP with $\psi=1$ (green nodes are usable, red nodes are damaged). (a) $\phi=0.45$, slightly below the site percolation threshold $\phi_c^{\text{site}}=1/2$~\cite{christensen2005complexity}. Damaged clusters are large. (b) $\phi=0.65$, above the standard percolation threshold. A GDC is present in this situation. (c) $\phi=0.85$, close to the birth of a GUC. Usable clusters are large but do not percolate, while a GDC still exists. (d) $\phi=0.95$, where a GUC is present. }
    \label{fig:triangular}
\end{figure}

In this section we present some additional results of numerical simulations for the neighbor-induced damage percolation in two-dimensional regular lattices. Some results of numerical simulations in various lattice topologies and various spatial dimensions $D$, for $\psi=1$, on what we called the usable threshold $\phi_c^{U}$ can be found in \cite{lobl2024efficient}, where the nonmonotonic behavior of the percolation threshold as a function of the spatial dimension $D$ is observed. Here we focus on $D=2$ and two lattice topologies: the square lattice (see Fig.~\ref{fig:square}) and the triangular lattice (see Fig.~\ref{fig:triangular}).

\begin{figure}
\center
\includegraphics[width=0.95\textwidth]{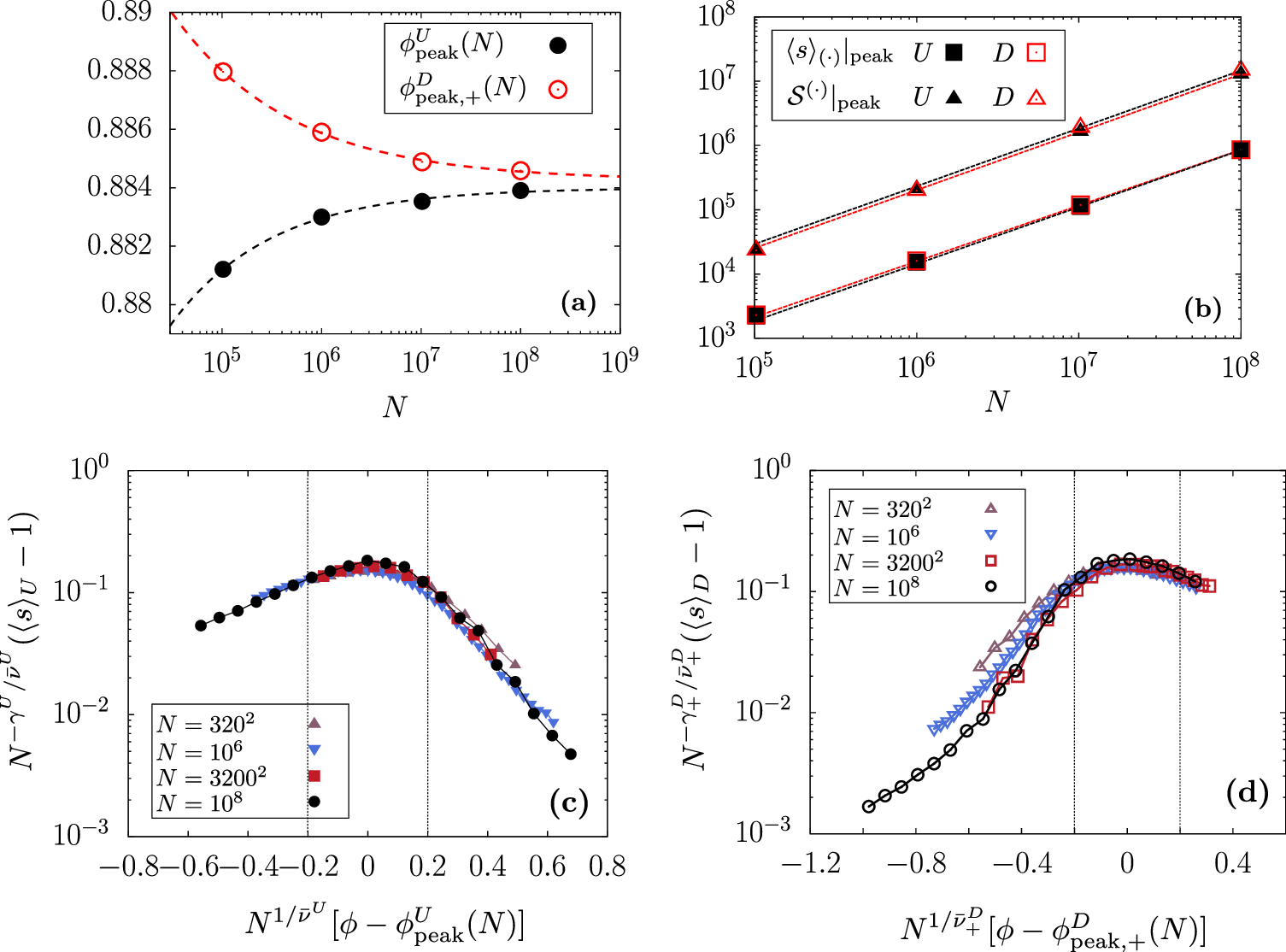}
    \caption{Finite-size scaling analysis of NIDP for $\psi=1$ in triangular lattices of various system sizes $N$. Results are averaged over $10^4$ realizations of the process. (a) The scaling of $\phi_{\text{peak}}^{U}(N)$ and $\phi_{\text{peak},+}^{D}(N)$ with $N$ as in Eq.~\eqref{eq:FSS_U} and \eqref{eq:FSS_D}. Dashed lines are fits with a function of the form $a+bN^{c}$. (b) The scaling of $ \langle s \rangle_{\text{peak}}$ and $\mathcal{S}_{\text{peak}}$ as in Eq.~\eqref{eq:FSS_U},\eqref{eq:FSS_D} and \eqref{eq:FSS_S_U},\eqref{eq:FSS_S_D}, respectively. Dashed lines are fit with a function of the form $ax^b$. (c) Collapse plot of $\langle s \rangle_{U}$, using the exact two-dimensional critical exponents $\gamma^{U}=\gamma_{D=2}=43/18$ and $\bar{\nu}^{U}=\bar{\nu}_{D=2}=8/3$. (d) As in (c) but for $\langle s \rangle_{D}$, with the same critical exponents $\gamma^{D}_{+}=\gamma_{D=2}$ and $\bar{\nu}^{D}_{+}=\bar{\nu}_{D=2}$.}
    \label{fig:FSS_triangular}
\end{figure}

Finite-size scaling analysis of the usable and damaged percolation phase transitions on the triangular lattice suggest that they are both described by the set of standard percolation critical exponents. In particular, we measure the position and the height of the peaks of average cluster size $\langle s \rangle_{U}$ and $\langle s \rangle_{D}$, which should scale as~\cite{cirigliano2024scaling}
\begin{align}
\label{eq:FSS_U}
&|\phi_c^{U}-\phi_{\text{peak}}^{U}(N)| \sim N^{-1/(\bar{\nu}^{U})}, \quad \left.\langle s \rangle_{U}\right|_{\text{peak}} \sim N^{\gamma^{U}/\bar{\nu}^{U}},\\
\label{eq:FSS_D}
&|\phi_+^{D}-\phi_{\text{peak},+}^{D}(N)| \sim N^{-1/(\bar{\nu}_{+}^{D})}, \quad \left.\langle s \rangle_{D}\right|_{\text{peak},+} \sim N^{\gamma^{U}/\bar{\nu}_{+}^{D}},
\end{align}
and the size of the largest usable ($\mathcal{S}^U$) and damaged ($\mathcal{S}^{D}$) clusters evaluated at the position of such peaks, which scale as~\cite{cirigliano2024scaling}
\begin{align}
\label{eq:FSS_S_U}
 \left.\mathcal{S}^U\right|_{\text{peak}} &\sim N^{\theta^{U}},\\
 \label{eq:FSS_S_D}
 \left. \mathcal{S}^{D}\right|_{\text{peak},+} &\sim N^{\theta^{D}_{+}},
\end{align}
where $\bar{\nu}=D \nu$, $\theta=1-\beta/\bar{\nu}$. We recall that the exact values of percolation critical exponents in $D=2$ are $\nu_{D=2}=4/3$, $\gamma_{D=2}=48/13$, $\beta_{D=2}=5/36$ (see the textbook~\cite{christensen2005complexity}). The two estimates of $\phi_c^{U}$ and $\phi_{+}^{D}$ from a fit in Fig.~\ref{fig:FSS_triangular}(a) give $\phi_c^{U}=0.8840(2)$, $\phi_{+}^{D}=0.8843(2)$, which are in agreement with the hypothesis $\phi_c^{U} \geq \phi_{+}^{D}$, implying that a GUC and a GDC cannot coexist in the triangular lattice. Further numerical studies are needed to better clarify this point. From the fit we also get estimates of the exponents $\bar{\nu}^U=2.3(4)$ and $\bar{\nu}^{D}_{+}=2.7(3)$, which are compatible with the exact known value of the critical exponent in $D=2$, i.e., $\bar{\nu}_{D=2}=8/3$. From the fits in Fig.~\ref{fig:FSS_triangular}(b) we get two estimates for the exponents $\gamma^{U}/\bar{\nu}^{U}=0.87(1)$ and $\gamma^{D}_{+}/\bar{\nu}^{D}_{+}=0.88(1)$, in good agreement with the exact value $\gamma_{D=2}/\bar{\nu_{D=2}}=43/48\simeq0.895$, and for the exponents $\theta^{U}=0.90(1)$ and $\theta^{D}_{+}=0.90(1)$, not far from the exact value $\theta_{D=2}=91/96 \simeq 0.948$. In Fig.~\ref{fig:FSS_triangular}(c),(d) a collapse plot of $N^{\gamma/\bar{\nu}} \langle s \rangle$ as a function of $N^{1/\bar{\nu}}[\phi-\phi_{\text{peak}}(N)]$ for the usable and damaged transition, respectively, with the exact critical exponents in $D=2$, shows a good compatibility with these known values. 

\bibliography{NIDP_BIB}